\documentclass[12pt,english,american]{article}
\usepackage[T1]{fontenc}
\usepackage[latin9]{inputenc}
\usepackage[letterpaper]{geometry}
\usepackage{array}
\usepackage{units}
\usepackage{multirow}
\usepackage{amsthm}
\usepackage{amsmath}
\usepackage{amssymb}
\usepackage{setspace}
\usepackage{esint}
\usepackage{dsfont}
\usepackage{color}
\usepackage{graphicx}

\usepackage[left, pagewise]{lineno}
\makeatletter

\providecommand{\tabularnewline}{\\}

\usepackage{enumitem}		% customizable list environments
\numberwithin{equation}{section}
\numberwithin{figure}{section}
\numberwithin{table}{section}

\theoremstyle{plain}
\newtheorem{thm}{\protect\theoremname}[section]
\newtheorem{remark}{Remark}[section]

\newtheorem{prop}{\protect\propositionname}[section]

\makeatother

\usepackage{babel}
  \addto\captionsamerican{\renewcommand{\lemmaname}{Lemma}}
  \addto\captionsamerican{\renewcommand{\propositionname}{Proposition}}
  \addto\captionsamerican{\renewcommand{\theoremname}{Theorem}}
  \addto\captionsenglish{\renewcommand{\lemmaname}{Lemma}}
  \addto\captionsenglish{\renewcommand{\propositionname}{Proposition}}
  \addto\captionsenglish{\renewcommand{\theoremname}{Theorem}}
   
\providecommand{\lemmaname}{Lemma}
\providecommand{\propositionname}{Proposition}
\providecommand{\theoremname}{Theorem}
\providecommand{\lemmaname}{Remark}

\newcommand\blfootnote[1]{%
  \begingroup
  \renewcommand\thefootnote{}\footnote{#1}%
  \addtocounter{footnote}{-1}%
  \endgroup
}

\begin{document}

\title{Parametric Estimation of Ordinary Differential Equations with Orthogonality
Conditions}

\author{{\large Nicolas J-B. Brunel$^1$, Quentin Clairon$^2$, Florence
d'Alch\'e-Buc$^{3,4}$}}

\date{May, 04 2013}

\maketitle
\begin{abstract}
Differential equations are commonly used to model dynamical deterministic systems in applications.  When statistical parameter estimation is required to calibrate theoretical models to data,  classical statistical estimators are often confronted to complex and potentially ill-posed optimization problem. As a consequence, alternative estimators to classical parametric estimators  are needed for obtaining reliable estimates. 
We propose a gradient matching approach for the estimation of parametric
Ordinary Differential Equations observed with noise. Starting
from a nonparametric proxy of a true solution of the ODE, we build
a parametric estimator based on a variational characterization of
the solution. As a Generalized Moment Estimator, our estimator
must satisfy a set of orthogonal conditions that are solved in the
least squares sense. Despite the use of a nonparametric estimator, we
prove the root-$n$ consistency and asymptotic normality of the Orthogonal
Conditions estimator. We can derive confidence sets thanks to a closed-form
expression for the asymptotic variance. 
Finally, the OC estimator is compared to classical estimators in several (simulated
and real) experiments and ODE models in order to show its versatility
and relevance with respect to classical Gradient Matching and Nonlinear
Least Squares estimators. In particular, we show on a real dataset
of influenza infection that the approach gives reliable estimates.
Moreover, we show that our approach can deal directly with more elaborated
models such as Delay Differential Equation (DDE).
% We obtain parameters and confidence sets without solving the DDE, and
% the values are consistent with the ones obtained with the ABC method.

\end{abstract}
\textbf{Key-words}: Gradient Matching,  Nonparametric statistics, Methods of Moments, Plug-in Property, Variational formulation, 
Sobolev Space.\blfootnote{
$^1$ ENSIIE \& Laboratoire Statistique et G\'enome, Universit\'e d'Evry Val d'Essonne, UMR CNRS 8071 - USC INRA   - FRANCE\\
$^2$ Laboratoire Analyse et Probabilit\'es, Universit\'e d'Evry Val d'Essonne - FRANCE \\
$^3$ Laboratoire IBISC, Universit\'e d'Evry Val d'Essonne - FRANCE \\
$^4$ INRIA-Saclay, LRI, Universit\'e Paris Sud, UMR CNRS 8623 - FRANCE \\}

\section{Introduction}

\subsection{Problem position and motivations}
Differential Equations are a standard mathematical framework for modeling
dynamics in physics, chemistry, biology, engineering sciences, etc and have proved their efficiency in describing
 the real world. Such models are defined thanks to a time-dependent vector field $\boldsymbol{f}$, defined on the state-space  $\mathcal{X}\subset\mathbb{R}^{d}$ 
 and that depends on a parameter $\theta\in\Theta\subset\mathbb{R}^{p}$,  $d,p\geq1$. The vector field is then a function from $[0,1] \times \mathcal{X} \times \Theta$ to $\mathbb{R}^{d}$. If $\phi(t)$ is the current state of the system, the time evolution is  given by the following  
Ordinary Differential Equation, defined for $t\in[0,1]$ by:
\begin{equation}
\dot{\phi}(t)=\boldsymbol{f}(t,\phi(t),\theta)\label{eq:ODE}
\end{equation}
where dot indicates derivative with respect to time. 
%is a time-dependent vector field from $\mathcal{X}\subset\mathbb{R}^{d}$
%to $\mathbb{R}^{d}$ which is parametrized by a parameter $\theta\in\Theta\subset\mathbb{R}^{p}$,  $d,p\geq1$.
An important task is then the estimation of the parameter $\theta$
from real data. \cite{Ramsay2007} proposed a significant improvement
to this statistical problem, and gave motivations for further statistical
studies. We are interested in the definition and in the optimality of
a statistical procedure for the estimation of the parameter $\theta$
from noisy observations $y_{1},\dots,y_{n}\in\mathbb{R}^{d}$ of a
solution at times $t_{1}<\dots<t_{n}$. 

Most works deal with Initial Value Problems (IVP), i.e. with ODE models having a given (possibly unknown) initial value $\phi(0)=\phi_{0}$. There exists then a unique solution $\phi(\cdot,\phi_{0},\theta)$
to the ODE (\ref{eq:ODE}) defined on the interval $\left[0,1\right]$, that depends smoothly on $\phi_{0}$ and $\theta$.

The estimation of $\theta$ is a classical problem of nonlinear
regression, where we regress $y$ on the time $t$.  If $\phi_{0}$ is known, the Nonlinear Least Square Estimator $\hat{\theta}^{NLS}$
(NLSE) is obtained by minimizing 
\begin{equation}
Q_{n}^{LS}(\theta)=\sum_{i=1}^{n}\vert y_{i}-\phi(t_{i},\phi_{0},\theta)\vert^{2}\label{eq:LSE}
\end{equation}
where $\left|\cdot\right|$ is the classical Euclidean norm. The NLSE,
Maximum Likelihood Estimator or more general M-estimators
\cite{vanGeer2000} are commonly used because of their good
statistical properties (root-$n$ consistency, asymptotic
efficiency), but they come with important computational
difficulties (repeated ODE integrations and presence of multiple local minima) that can decrease their interest. We refer to \cite{Ramsay2007} for a detailed overview of the previous works in this field. An adapted NLS estimator (dedicated the specific difficulties of ODEs) is also introduced and studied in \cite{XueMiaoWuaos2010}. 

Global optimization methods are then often used,
such as simulated annealing, evolutionary algorithms (\cite{Moles2003}
for a comparison of such methods). 
Other classical estimators are obtained by interpreting noisy ODEs as state-space models: 
filtering and smoothing technics can be used for parameter
inference \cite{CaMoRy05}, which can provide estimates with reduced computational complexity \cite{Quach2007,Ionides06,Ionides2011}. 

Nevertheless, the difficulty of the optimization problem is the outward
sign of the illposedness of the inverse problem of ODE parameter estimation,
\cite{Engl2009}. Hence some improvements  on classical
estimation have been proposed by regularizing the statistical inference in an appropriate
way. 

Starting from different methods used for solving ODEs, different estimators can be developed based on a mixture of nonparametric estimation and collocation approximation. 
This gives rise to Gradient Matching (or Two-Step) estimators that consists in approximating
the solution $\phi$ with a basis expansion, such as cubic splines. The rationale is to
estimate nonparametrically the solution $\phi$ by $\hat{\phi}=\sum_{k=1}^{L}\hat{c}_{k}B_{k}$
so that we can also estimate the derivative $\dot{\hat{\phi}}$. An
estimator of $\theta$ can be obtained  by looking for the parameter
that makes $\hat{\phi}$ satisfy the differential equation (\ref{eq:ODE})
in the best possible manner. Two different methods have been proposed, based on a $L^{2}$
distance between $\dot{\hat{\phi}}$ and $\boldsymbol{f}(t,\hat{\phi},\theta)$: 
The first one, called the {\it two-step method}, was originally proposed by \cite{Varah1982}, and has
been particularly developed in (bio)chemical engineering \cite{Madar2003,Voit2004,Poyton2006}.
It avoids the numerical integration of the ODE and usually gives rise
to simple optimization program and fast procedures that usually performs
well in practice. The statistical properties of this two stage estimator
(and several variants) have been studied in order to understand the
influence of nonparametric technics to estimate a finite dimensional
parameter \cite{Brunel2008,LiangWujasa2008,GugushviliKlaassen2010}.
While keeping the same kind of numerical approximation of the solution,
\cite{Ramsay2007} proposed a second method based on the generalized smoothing approach for determining
at the same time the parameter $\theta$ and the nonparametric estimation
$\hat{\phi}$. The essential difference between these two approaches
is that the nonparametric estimator in the generalized smoothing approach
is computed adaptively with respect to the parametric model, whereas
two-step estimators are ``model-free smoothing''. 

We introduce here a new estimator that can be seen as an improvement and a generalization
of the previous two-step estimators. It uses also a nonparametric
proxy $\hat{\phi}$, but we modify the criterion used to identify
the ODE parameter (i.e. the second step). The initial motivations are 
\begin{itemize}
\item to get a closed-form expression for the asymptotic variance and confidence sets, 
\item to reduce sensitivity to the estimation of the derivative in Gradient Matching approaches, 
\item to take into account explicitly time-dependent vector field, with potential discontinuities in time. 
\end{itemize}
The most notable feature of the proposed method is the use of a variational formulation of the differential equations instead
of the classical point-wise one, in order to generate conditions to satisfy. This formulation is rather general and can cover
a greater number of situations: we come up with a generic class of estimator of Differential Equations (e.g Ordinary, Delay, Partial, Differential-Algebraic), that can incorporate relatively easily prior knowledge about the true solution. In addition to the versatility of the method, the criterion is built in order to offer  computational tractability, that implies that we can give a precise description of the asymptotics
and give the bias and variance of the estimator.
We also give a way to ameliorate adaptively our estimator and to compute asymptotic
confidence intervals. 

First, we introduce the statistical ODE-based model and main assumptions,
we motivate and describe our estimator, and show its consistency. Then, we provide
a detailed description of the asymptotics, by proving its root-$n$
consistency and asymptotic normality. Based on the asymptotic approximation,
we give a closed-form expression of the asymptotic variance, and we
address the problem of obtaining the best variance through the choice
of an appropriate weighting matrix. Finally, we provide some insights
into the practical behavior of the estimator through simulations and by considering two real-data examples.
The objective of the experiments parts is to show the interest of OC with
respect to the nonlinear least squares and classical gradient matching
estimators. 

\subsection{Examples}
We motivate our work in detail by presenting two models that are relatively common and simple but that nevertheless causes difficulty for estimation. 
\subsubsection{Ricatti ODE\label{sec:RicattiEx}}
The (scalar) Ricatti equation is defined by a quadratic vector field $f(t,x)=a(t) x^2+ b(t) x + c(t) $ where $a(\cdot), b(\cdot), c(\cdot)$ are time-varying functions. This equation arises naturally in control theory for solving linear-quadratic control problem \cite{Sontag1998}, or  in mathematical finance, in the analysis of stochastic interest rate models \cite{BrigoMercurio}. We consider one of the simplest case where $a$ is constant, $b=0$ and $c(t)=c\sqrt{t}$. The objective is to estimate parameters $a,c$ from the noisy observations $y_i = \phi(t_i)+\epsilon_i$ for $t_i \in [0,14]$. Here the true parameters are $a=0.11$, $c=0.09$ and $\phi_0=-1$, and one can see the solution and simulated observations in figure \ref{fig:Ricatti_Solutions}. Although the solution $\phi$ is smooth in the parameters, there  exists no closed form and simulations are required for implementing NLS and classical approaches. The hard part in this equation is due to the extreme sensitivity of the squared term in the vector field: for small differences in the parameters or initial condition, the solution can explode before reaching the final time $T=14$ \ref{fig:Ricatti_Solutions}. Explosions are not due to numerical instability but to the failure of (theoretical) existence of a global solution on the entire interval (e.g the \emph{tangent} function is solution of $\dot{\phi}=\phi^2 +1$, $\phi(0)=0$ and explodes at $t=\frac{\pi}{2}$). The explosions have to be handled in estimation algorithms and this slows down the exploration of the parameter space (which can be difficult for high-dimensional state or parameter spaces). 
Nevertheless, we show in the experiment part that NLS or Gradient Matching can do well for parameter estimation, but some additional difficulties  does appear when the time-dependent function $c(\cdot)$ has abrupt changes. We consider the case where $c(t)=c\sqrt{t} - d' \mathds{1}_{[T_r, T]}$, $T_r$ is a change-point time, with $d'>0$. This situation is classical (e.g in engineering) where some input variables $t\mapsto u(t)$ modify the evolution of the system $\dot{\phi}=f(t,\phi(t)) +u(t)$ (typically it can be the introduction of a new chemical species in a reactor at time $T_r$), see figure \ref{fig:Ricatti_Solutions}. The Cauchy-Lipschitz theory for existence and uniqueness of solutions to time-discontinuous ODE is extended straightforwardly with measure theoretic arguments \cite{Sontag1998}. The (generalized) solution is defined almost everywhere and belongs to a Sobolev space. For sake of completeness, we provide a generalized version of the Cauchy-Lipschitz theorem for IVP in \emph{Supplementary Material I}.  This abrupt change causes some difficulties in estimating non-parametrically the solution and its derivative, which can make Gradient Matching less precise. We consider then the estimation of the two additional parameters $d'$ and $T_r$. Hence, the parameter estimation problem can be seen as a change-point detection problem, where the solution $\phi$ still depends smoothly in the parameters. Nevertheless, in the case of the joint estimation of $a,c, d'$ and $T_r$, the particular influence of the parameter $T_r$ makes the problem much more difficult to deal with for classical approaches as it is suggested by the objective functions in \emph{Supplementary Material II}. The variational formulation for model estimation gives a seamless approach for estimating models which possess time discontinuities.
%FIGURES (OBJECTIVE FUNCTION FOR NLS AND TWO STEP - asymptotic or not).  

\begin{center}
\begin{figure}[h]
\begin{centering}
\begin{tabular}{|c|c|}
\hline 
\includegraphics[scale=0.3]{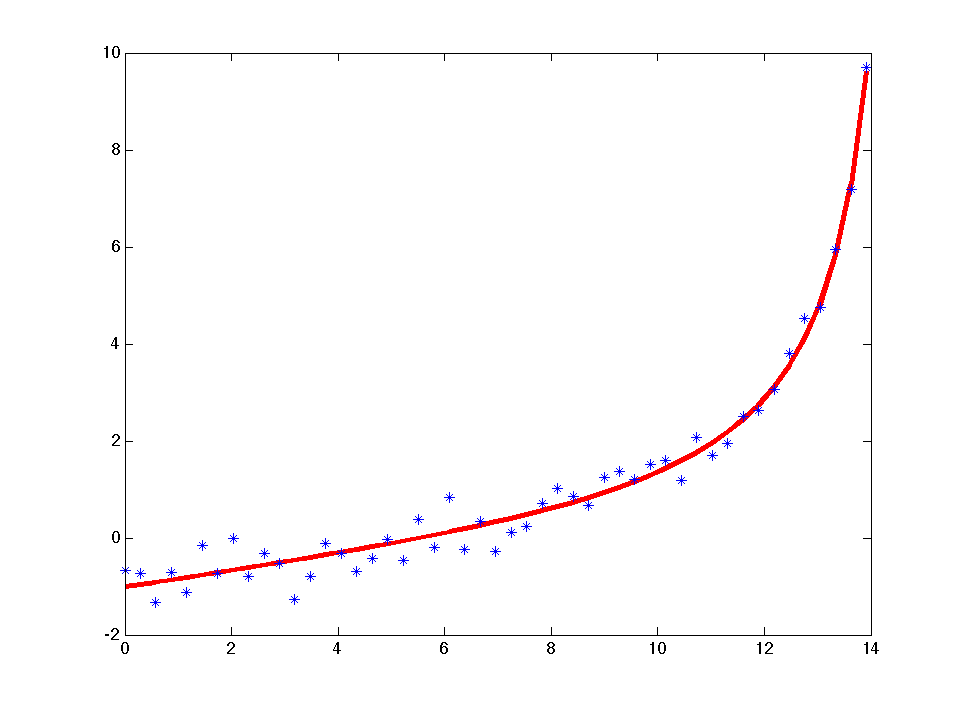} & \includegraphics[scale=0.3]{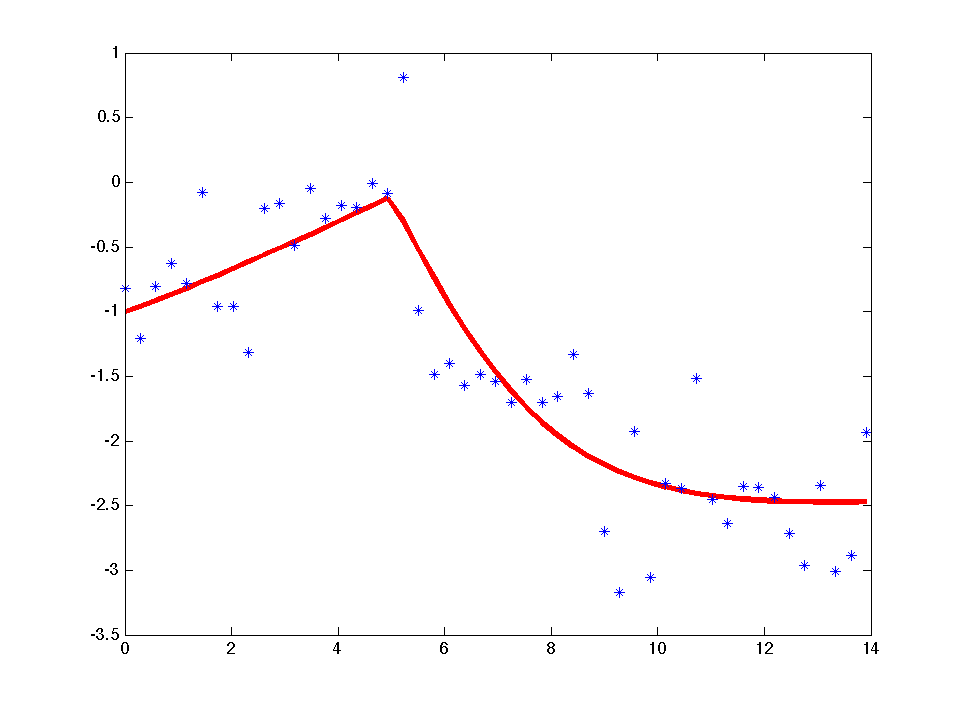}\tabularnewline
\hline 
\end{tabular}
\par\end{centering}
\caption{Solutions of Ricatti ODE with noisy observations ($n=50$, $\sigma=0.4$).
Left figure is smooth time-dependent ODE. Right figure has a change-point at time $T_{r}=5$ ($d'=1$). \label{fig:Ricatti_Solutions}}
\end{figure}
\par\end{center}

\subsubsection{Dynamics of Blowfly populations\label{sec:BlowflyEx}}
\begin{center}
\begin{figure}
\begin{centering}
\includegraphics[scale=0.5]{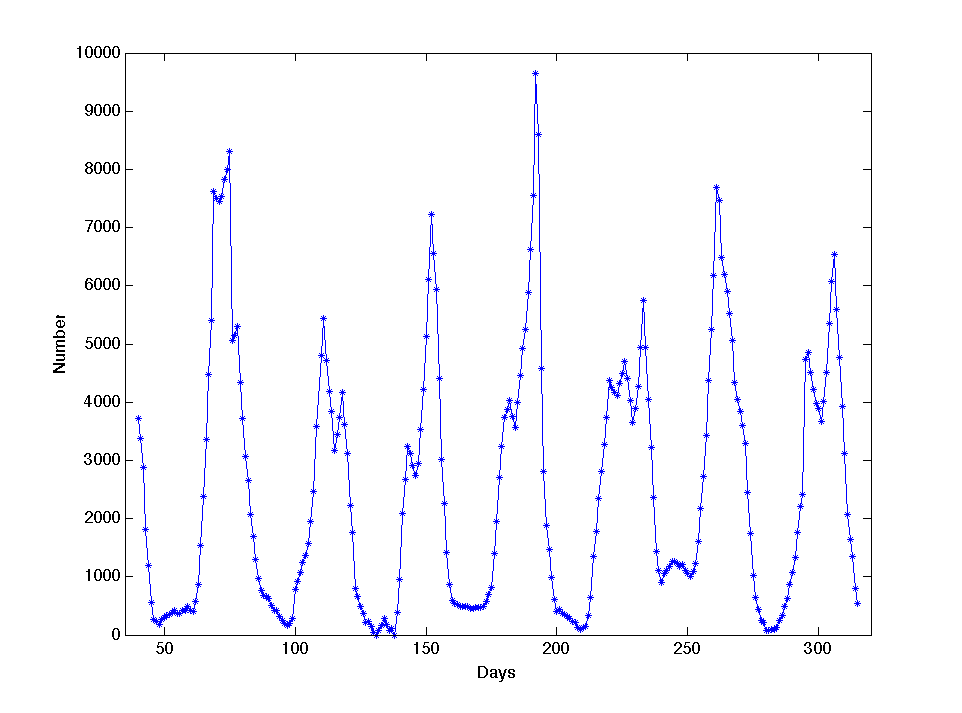}
\par\end{centering}
\caption{Blowfly Data, collected by Nicholson \label{fig:BlowflySeries}}
\end{figure}
\end{center}

The modeling of the dynamics of population is a classical topic in ecology an more generally in biology. Differential Equations can describe very precisely the mechanics of evolution, with birth, death and migration effects. The case of single-species models is the easiest case to consider, as interactions with rest of the world can be limited, and the acquisition of reliable data is easier. In the 50s, Nicholson measured quite precisely the dynamics of a blowfly population, known as  Nicholson's experiments \cite{Nicholson1957}. The data are relatively hard to model, and it is common to use Delay Differential Equation (DDE) whose general form is $\dot{N}(t)=f \left( N(t), N(t-\tau), \theta \right)$, in order to account for the almost chaotic behavior of  the data, see figure \ref{fig:BlowflySeries}.
Nicholson's dataset is now a classical benchmark for evaluating time series algorithms due its intrinsic complexity. Nevertheless, the following DDE is commonly acknowledged as a correct model \cite{ellner2006dynamic, Murray2004}:
\begin{equation}
\dot{N}= P N(t-\tau) \exp \left( -N(t-\tau)/N_0 \right) -\delta N(t) \label{eq:NicholsonDDE}\end{equation} whose parameter fitting (of $P, N_0, \delta$) remains delicate. In particular classical NLS are difficult to use in this setting as the initial condition, which is a function defined on $[-\tau, 0]$, is unknown. Alternative solutions, such as Gradient Matching or Bayesian Methods (based on ABC, \cite{wood2010statistical}) give reliable estimates that reproduce the observed dynamics without estimation of the initial condition. 
These aforementioned methods use particular statistics or functions of the model that provides high-level information on the parameters. The Orthogonal Conditions estimator has a similar approach for dealing with the estimation of Differential Equations. 

\section{Differential Equation Model and Gradient Matching \label{sec:Differential-Equation-Model}}

\subsection{ODE models and Gradient Matching}
For ease of readability, we focus on a two-dimensional system of ODEs. In our case, as there is no computational and theoretical differences between the situation $d=2$ and $d>2$, there is no lack of generality by this assumption. 
We consider noisy observations  $Y_{1},\dots,Y_{n}\in\mathbb{R}^{2}$ 
of the function $\phi^{*}$ measured at random times $t_{1}<\dots<t_{n}\in\left[0,1\right]$: 
\begin{equation}
Y_{i}=\phi^{*}(t_{i})+\epsilon_{i}\label{eq:Observations}
\end{equation}
where $\epsilon_{1},\dots,\epsilon_{n}$ are i.i.d with $E(\epsilon_{i})=0$ and $V(\epsilon_{i})=\sigma^{2}I_{2}$. We suppose that the regression
function $\phi^{*}$ belongs to the Sobolev space $H^{1}=\left\{ u\in L^{2}(\left[0,1\right])\right.$
$\left|\dot{u}\in L^{2}(\left[0,1\right])\right\} $, and $\phi^{*}$
is a solution to the parametrized Ordinary Differential
Equation (\ref{eq:ODE}), i.e.  there exists a \emph{true} parameter $\theta^{*}\in\Theta\subset\mathbb{R}^{p}$
such that for $t\in\left[0,1\right]$ almost everywhere (a.e.) 
\begin{equation}
\dot{\phi^{*}}(t)=\boldsymbol{f}\left(t,\phi^{*}(t),\theta^{*}\right)\label{eq:TrueODE}
\end{equation}
where $\boldsymbol{f}=(f_1, f_2)$ is a vector field from $\left[0,1\right]\times\mathcal{X}\times\Theta$
to $\mathbb{R}^{2}$, where $\mathcal{X}\subset\mathbb{R}^{2}$.

The statistical problem can be seen as a noisy version of a parametrized Multipoint Boundary-Value Problem (MBVP, \cite{BellKagKal62}). MBVP deals with the existence, uniqueness and computation of a solution $\phi^*$ to equation 
 (\ref{eq:ODE}), with general boundary conditions $\phi^*(t_1)=y_1, \dots, \phi^*(t_n)=y_n,\, n\geq2$. Obviously, MBVP is a much more difficult problem than the classical Initial Value Problem although some theoretical results do exist in some restricted cases (\cite{Bellman66, OjiWel79} and references therein). On the computational side, numerous algorithms such as collocation, multiple shooting,... have been proposed to solve general Boundary Value Problems, \cite{ascher1988numerical}. Among them, the 2 points Boundary Value Problem (BVP) where $G\left( \phi^*(0),\phi^*(1)\right)=0$ with  $G$ a given function, is one of the most common and important one, as it arises in numerous applications (physics, control theory,\dots). We emphasize that a convenient way to deal theoretically and computationally with BVP, in particular linear second order differential ODEs, is not based on an adaptation of  the IVP theory, but it rather involves elaborated concepts from functional analysis such as  weak derivative, variational formulation and Sobolev spaces \cite{conway1990course}.
If we denote the inner product of $L^{2}$ as $\forall\varphi,\psi\in L^{2}\left(\left[0,1\right]\right),\:\left\langle \varphi,\psi\right\rangle =\int_{0}^{1}\varphi(t)\psi(t)dt$,
the weak derivative of the function $g$ in $H^{1}$ is not defined
point-wise but as the function $\dot{g}\in L^{2}$ satisfying $\left\langle \dot{g},\varphi\right\rangle =-\left\langle g,\dot{\varphi}\right\rangle $,
for all function $\varphi$ in $C^{1}$ with support included in $\left]0,1\right[$
(denoted $C_{C}^{1}\left(\left]0,1\right[\right)$). Of course, if
$t\mapsto\phi\left(t,x_{0},\theta\right)$ is a $C^{1}$ function
on $\left]0,1\right[$, the classical derivative $\dot{\phi}$ is
also the weak derivative. 
We introduce then the (weak) variational formulation of
the ODE (\ref{eq:ODE}): a weak solution $g$ to (\ref{eq:ODE})
is a function in $H^1$ such that $\forall\varphi\in C_{C}^{1}\left(\left]0,1\right[\right)$
\begin{equation}
\int_{0}^{1}\boldsymbol{f}(t,g(t),\theta)\varphi(t)dt+\int_{0}^{1}g(t)\dot{\varphi}(t)dt=0\label{eq:VariationalFormulation}
\end{equation}
This variational formulation is the key of the Finite Elements Method which is the reference approach for solving Boundary Value Problems and Partial Differential Equations, \cite{Brenner2008}.  In the case of ODEs, this formulation is not well used for computing solutions, because the geometry of the (1-D) interval $]0,1[$ is simple, and it is easy to build a spline approximation by collocation that solves approximately the ODE.
Nevertheless, the characterization (\ref{eq:VariationalFormulation}) is useful for the statistical inference task, as it enables to give necessary conditions for a good estimate. In particular, we emphasize that we do not solve the ODE, but we want to identify a parameter $\theta$ indexing the vector field $\boldsymbol{f}$. Hence, we develop a new algorithmic approach, different from the one used for solving the direct problem. 

\subsection{Definition}

We define a new gradient matching estimator based on (\ref{eq:VariationalFormulation}):
starting from a nonparametric estimator $\hat{\phi}$, computed from
the observations $(t_{i},y_{i}),\: i=1,\dots,n$,  we want to find
the parameter $\theta$ that minimizes the discrepancy between the
parametric derivative $t\mapsto\boldsymbol{f}\left(t,\hat{\phi}(t),\theta\right)$
and a nonparametric estimate of the derivative, e.g. $\dot{\hat{\phi}}$.
A classical discrepancy measure is the $L^{2}$ distance, that gives
rise to the two-step estimator $\hat{\theta}^{TS}$ defined as $\hat{\theta}^{TS}=\arg\min_{\theta\in\Theta}R_{n,w}(\theta)$
where \begin{equation}
R_{n,w}(\theta)=  \int_{0}^{1} \vert \dot{\hat{\phi}}(t)-\boldsymbol{f}\left(t,\hat{\phi}(t),\theta\right)\vert^{2}w(t)dt. \label{eq:ClassicalTS}
\end{equation}
This estimator is consistent for several usual nonparametric estimators
\cite{Brunel2008,LiangWujasa2008,GugushviliKlaassen2010}, but
the use of a positive weight function $w$ vanishing at the boundaries
($w(0)=w(1)=0$) is needed to get the classical parametric root-$n$
rate. The importance of the weight function $w$ for the asymptotics of  $\hat{\theta}^{TS}$  is assessed by theorem 3.1 in \cite{Brunel2008}. Indeed, if $w$ does not vanish at the boundaries, then $\hat{\theta}^{TS}$ does not have  a root-$n$ rate, because the asymptotics is then dominated by the nonparametric estimates $\hat{\phi}(0)$ and $\hat{\phi}(1)$. 
The usefulness of such weighting function is well acknowledged in nonparametric or  semiparametric estimation. For instance,  the so-called weighted average derivative is based on a similar weight function in order to get estimators with parametric rate in partial index models \cite{NewSto93}. \\
The use of a nonparametric proxy (instead of a solution to be computed) gives  the opportunity to consider parameter estimation in $f_1$ and in $f_2$ separately. For this reason and ease of readability, we consider only the estimation of the parameter $\theta_1$ 
when $\boldsymbol{f}$ can be written $\boldsymbol{f}(t,x,\theta)=\left(f_{1}(t,x,\theta_{1}),f_{2}(t,x,\theta_{2})\right)^{\top}$ and
 $\theta=\left(\theta_{1},\theta_{2}\right)^{\top}$ ($\theta_{i}\in\mathbb{R}^{p_{i}}$
and $p_{1}+p_{2}=p$). The joint estimation of $\theta=\left(\theta_{1},\theta_{2}\right)^{\top}$ can be done by stacking the observations into a single column: there is no consequence on the asymptotics, but the estimator covariance matrix has to be slightly modified in order to take into account the correlations between the two equations $f_1$ and $f_2$. 
Having said that, we write simply $f=f_1$ and $\theta=\theta_{1}$ and we consider only one equation $\dot{x}_1=f(t,x,\theta)$. We use a nonparametric estimator $\hat{\phi}=(\hat{\phi}_1, \hat{\phi}_2)$
of $\phi^{*}:\left[0,1\right] \rightarrow \mathbb{R}^2$. 

Starting from (\ref{eq:VariationalFormulation}), a reasonable estimator $\hat{\theta}$ should satisfy the weak formulation 
\begin{equation}
\forall\varphi\in C_{C}^{1}\left(\left]0,1\right[\right),\:\int_{0}^{1}f\left(t,\hat{\phi}(t),\hat{\theta}\right)\varphi(t)dt+\int_{0}^{1}\hat{\phi}_1(t)\dot{\varphi}(t)dt=0.\label{eq:InfOrthoCond}
\end{equation}
The vector space $ C_{C}^{1}\left(\left]0,1\right[\right)$ is not tractable for variational formulation, and one prefers Hilbert space with a structure related to $L^2$. In our case, we use $H^1_0 = \{h \in H^1 \vert h(0)=h(1)=0\}$ which has a simple description within $L^2$: an orthonormal basis is given by the sine functions $t \mapsto \sqrt{2}\sin(\ell \pi t), \ell\geq1$ and we have 
\begin{equation}
H_{0}^{1}=\left\{ \sum_{\ell=1}^{\infty}a_{\ell}\sqrt{2}\sin\left(\ell\pi t\right)\left|\sum_{\ell=1}^{\infty}\ell^{2}a_{\ell}^{2}<\infty\right.\right\} \label{eq:H_01}
\end{equation}
%As a consequence, the condition (\ref{eq:InfOrthoCond}) is replaced by the simpler condition based on countable set of orthogonal
%conditions: 
Hence, it suffices to consider a countable number of orthogonal conditions  (\ref{eq:InfOrthoCond})  defined, for instance, with the test functions $\varphi_{\ell}=\sqrt{2}\sin(\ell \pi t)$, $\forall\ell\geq1$: % For all $\ell \leq 1$, 
%\begin{equation}
%\forall\ell\geq1,\:\int_{0}^{1}f\left(t,\hat{\phi}(t),\hat{\theta}\right)\varphi_{\ell}(t)dt+\int_{0}^{1}\hat{\phi}_{1}(t)\dot{\varphi}_{\ell}(t)dt=0.\label{eq:InfOrthoCond-1}
%\end{equation}
\begin{equation}
\mathcal{C}_\ell(\theta): \: \int_{0}^{1}f\left(t,\hat{\phi}(t),\hat{\theta}\right)\varphi_\ell(t)dt+\int_{0}^{1}\hat{\phi}(t)\dot{\varphi_\ell}(t)dt=0.\label{eq:FiniteInfOrthoCond}
\end{equation}
More generally, we consider a family of orthonormal functions $\varphi_\ell \in H_0^1$, with $\ell \geq 1$, and we introduce the vector space $\mathcal{F}= \overline{span \{\varphi_\ell, \ell \geq 1\}}$. The vector space $\mathcal{F}$ may not be necessarily dense in $H_0^1$, as the functions $\varphi_\ell$ could be chosen for computational tractability or because of a natural interpretation (for instance B-splines, polynomials, wavelets, ad-hoc functions, \dots).  For this reason, we introduce the orthogonal decomposition of $H_0^1=\mathcal{F}\oplus\mathcal{F}^{\bot}$, where $\mathcal{F}^{\bot}=\{g\in H_0^1 \vert \langle g, \varphi \rangle=0, \varphi \in \mathcal{F} \}$, and we can have $\mathcal{F}\neq H_0^1$. 
% But if $\mathcal{F}$ is not dense $H_0^1$, then the condition (\ref{eq:InfOrthoCond-1}) is less informative about the model, and this could imply a loss for parameter estimation.
In general, an estimator $\hat{\theta}$ satisfying $\mathcal{C}_\ell(\hat{\theta})$ for $\ell \geq 1$ also approximately satisfies (\ref{eq:InfOrthoCond}). However in practice, we will use a finite set of orthogonal constraints defined by $L$ test functions ($L>p$).  \\
In order to discuss the influence of the choice of $\mathcal{F}$ and of finite dimensional subspace  spanned by $\varphi_{1},\dots,\varphi_{L}$ we introduce the nonlinear operator $\mathcal{E}\,:\,(g,\theta)\mapsto\mathcal{E}\left(g,\theta\right)$,
such that $t\mapsto\mathcal{E}\left(g,\theta\right)(t)=f\left(t,g(t),\theta\right)$. 
%Nevertheless, we will relate the statistical properties of the estimator to the test functions $\left(\varphi_{\ell}\right)_{1\leq\ell\leq L}$ in order to give a practical way to choose them.
%whose property relies on the regularity of $\mathcal{E}$ in $g$
%and $\theta$, the latter being highly dependent on the regularity
%of $f$ on $\mathcal{D}\times\Theta$. 

For all $\theta$ in $\Theta$ and $g$ in $H^{1}$, the Fourier coefficients of $\mathcal{E}(g,\theta)-\dot{g}$ in the basis $\left(\varphi_{\ell}\right)_{\ell\geq1}$ are $e_{\ell}\left(g,\theta\right) = $ $ \left\langle \mathcal{E}(g,\theta)-\dot{g},\varphi_{\ell}\right\rangle= $ $ \left\langle \mathcal{E}(g,\theta),\varphi_{\ell}\right\rangle +\left\langle g,\dot{\varphi}_{\ell}\right\rangle$, and we introduce the vectors in $\mathbb{R}^{L}$
$\boldsymbol{e}_{L}(g,\theta)=\left(e_{\ell}(g,\theta)\right)_{\ell=1..L}$
and $\boldsymbol{e}_{L}^{*}(\theta)=\left(e_{\ell}(\phi^{*},\theta) \right)_{\ell=1..L}$. Finally, our estimator is defined by the minimization of the quadratic form
$Q_{n,L}(\theta)=\left|\boldsymbol{e}_{L}(\hat{\phi},\theta)\right|^{2}$:
\begin{equation}
\hat{\theta}_{n,L}=\arg\min_{\theta\in\Theta}Q_{n,L}(\theta).\label{eq:OrthogonalEstimator}
\end{equation}
$\hat{\theta}_{n,L}$ is the parameter that ``almost'' vanishes
the first $L$ Fourier coefficients in the orthogonal decomposition
of $H_0^1=\mathcal{F}\oplus\mathcal{F}^{\bot}$:
\begin{eqnarray*}
\mathcal{E}(g,\theta)-\dot{g} & = & \boldsymbol{E}{}_{L}(g,\theta)+\boldsymbol{R}{}_{L}(g,\theta)+\boldsymbol{E}_{\mathcal{F}}^{\bot}\left(g,\theta\right)
\end{eqnarray*}
with $\boldsymbol{E}{}_{L}(g,\theta)=\sum_{\ell=1}^{L}e_{\ell}\left(g,\theta\right)\varphi_{\ell}$,
$\boldsymbol{R}{}_{L}(g,\theta)=\sum_{\ell>L}e_{\ell}\left(g,\theta\right)\varphi_{\ell}$
and $\boldsymbol{E}_{\mathcal{F}}^{\bot}\left(g,\theta\right)\in\mathcal{F}^{\bot}$. 

The function $\boldsymbol{E}_{\mathcal{F}}^{\bot}\left(\phi^{*},\theta\right)$
represents the behavior of $\mathcal{E}(g,\theta)-\dot{g}$ at the
boundaries of the interval $\left[0,1\right]$. As $\hat{\phi}$ approaches
$\phi^{*}$ asymptotically in supremum norm, the objective function
$Q_{n,L}(\theta)$ is close to $Q_{L}^{*}(\theta)=\left\Vert \boldsymbol{E}{}_{L}(\phi^{*},\theta)\right\Vert _{L^{2}}^{2}$. The discriminative power of $Q_{L}^{*}(\theta)$ can be analyzed locally around its global minimum $\theta^*_L$, as it behaves approximately as the quadratic form $Q_{L}^{*}(\theta)\approx\left(\theta-\theta^{*}_L\right)^{\top}\mathbf{J}_{\theta,L}^{*\top}\mathbf{J}_{\theta,L}^{*}\left(\theta-\theta^{*}_L\right)$
where $\mathbf{J}_{\theta,L}^{*}$ is the matrix in $\mathbb{R}^{L\times p}$
with entries $\int_{0}^{1}f_{\theta_{j}}(t,\phi^{*}(t),\theta^{*}_L)\varphi_{\ell}(t)dt$,
for $j=1,\dots,p$, $\ell=1,\dots,L$. 

\subsection{Boundary Conditions and Construction of Orthogonal Conditions}

The construction of the orthogonal conditions $e_\ell(\theta)$ exposed in the previous section is generic and can be proposed for numerous types of Differential Equations, in particular for Ordinary and Delay Differential Equations. Moreover,  similar orthogonal conditions  could be also derived for solutions of PDEs with a relevant set of test functions $\varphi$, but this extension is  beyond the scope of the present paper.
A process for deriving "regular" orthogonal conditions, (i.e that gives rise to root-$n$ consistent estimator, as it is shown in section \ref{sec:Asymptotics}) is to use conditions $\mathcal{C}_\ell(\theta)$ with an integral expression $\int_0^1 h_\ell \left( t,\hat{\phi}(t),\theta \right) dt$. The function $h_\ell : (t,x,\theta) \longrightarrow \mathbb{R} $ must be smooth and must satisfy the remarkable identity  $\int_0^1 h_\ell \left( t,\phi^*(t),\theta^*  \right)=0 $. The variational formulation generates functions $h_\ell(t,x,\theta)=\left( f\left( t , x ,\theta \right)\varphi_\ell(t)  -  \dot{\varphi}_\ell(t) x \right ) $ whereas the classical Gradient Matching considers a single function $h(t,x,y,\theta)=\left\Vert f\left( t , x ,\theta \right) - y \right\Vert ^2 \varphi (t) $, and the variable $y$ is evaluated along the derivative $\dot{\phi}(t)$. The asymptotic analysis shows that the dependency in $y$ can be removed and that $h'$ behaves in fact as a function $h(t,x,\theta)$. \\
The OC framework then generalizes the classical TS estimator and gives ways to ameliorate it. Among other, the use of the boundary vanishing function  $\varphi$ implies an information loss close to the boundaries. This loss  can be sensible in terms of estimation quality, and should be avoided when the boundary values are known. For instance, for an IVP with known initial condition $\phi(0)=\phi_0$, we can derive an orthogonal condition that  takes into account the knowledge of $\phi_0$. By direct computation, we have \begin{eqnarray*}
\int_{0}^{1}h(t,\phi(t),\theta)dt & = & \int_{0}^{1}f(t,\phi(t),\theta)\varphi(t)dt-\left[\phi(1)\varphi(1)-\phi(0)\varphi(0)\right]\\
 &  & +\int_{0}^{1}\phi(t)\dot{\varphi}(t)dt.
\end{eqnarray*}
If $\phi(1)$ is unknown, but $\phi(0)$ is known, it suffices to take $\varphi$ such that $\varphi(1)=0$ and $\varphi(0)\neq0$. The orthogonal condition still have the same expression $h(t,x,\theta)$. The same adaptation can be done when boundary values of the derivative are known (called Neumann's condition), for instance $\dot{\phi}(1)=\phi_1'$ is known. Indeed, the ODE gives a relationship between the second order derivative $\ddot{\phi}$ and the state $\phi$, as  $\ddot{\phi}(t)=\partial_xf(t,\phi,\theta)f(t,\phi,\theta)$. By choosing $\varphi$ such that $\varphi(0)=0$ and by Integration By Part, the following identity
 \begin{eqnarray*}
\varphi(1)\phi_1' = \int_{0}^{1}\partial_xf(t,\phi,\theta)f(t,\phi,\theta)\varphi(t)dt +\int_{0}^{1}f(t,\phi,\theta)\dot{\varphi}(t)dt
\end{eqnarray*}
gives a new condition that exploits the behavior of the solution at the boundary. Obviously,  these conditions can be successfully used if the nonparametric proxy satisfies the boundary conditions of interest. At the contrary, it seems rather difficult to integrate such information about the boundary within the criterion $R_{n,w}(\theta)$.
The orthogonal conditions introduced in the previous section are a direct exploitation of the ODE model, and the introduction of the space $\mathcal{F}$ is  a way to deal with the problem of the choice of the number of conditions and their type. Nevertheless, it would be useful to introduce model specific conditions $h(t,\phi(t),\theta)$ which are known to have a vanishing integral for $\theta=\theta^*$.  
Our estimator can be thought as a Generalized Method of Moments estimator, but where Moments do characterize curves and not probability distributions. A similar idea has been developed recently in the context of functional data analysis \cite{James2007}. 

\section{Consistency of the Orthogonal Conditions estimator\label{sec:Definition-and-Consistency}}

In order to obtain precise results with closed-form expression for
the bias and variance estimators, we consider series estimators,
i.e. estimators expressed as $\hat{\phi}_{j}=\sum_{k=1}^{K}\hat{c}_{k,j}p_{kK}=\hat{\boldsymbol{c}}_{j}\boldsymbol{p}^{K}$,
where $\boldsymbol{p}^{K}=\left(p_{1K}\right.,$\,\,\,$\dots,$\,$\left.p_{kK}\right)$
is a vector of approximating functions and the coefficients $\hat{\mathbf{c}}_{j}=(\hat{c}_{k,j})_{k=1..K}$
are computed by least squares. For notational simplicity, we use the
same functions (and the same number $K$) for estimating $\phi_{1}^{*}$
and $\phi_{2}^{*}$. We denote $P^{K}=\left(p_{kK}(t_{i})\right)_{1\leq i,k\leq n,K}$
the design matrix and $\mathbf{Y}_{j}=\left(y_{i,j}\right)_{i=1..n}$
the vectors of observations. Hence, the estimated coefficients $\hat{\mathbf{c}}_{j}=\left(P^{K\top}P^{K}\right)^{\dagger}P^{K\top}\mathbf{Y}_{j}$
(where $\dagger$ denotes a generalized inverse) gives rise to the
so-called hat matrix $H=P^{K}\left(P^{K\top}P^{K}\right)^{\dagger}P^{K\top}$
and the vector of smoothed observations is $\hat{\boldsymbol{\phi}}_{j}=H\mathbf{Y}_{j}$,
$j=1,2$. One can typically think of regression splines, \cite{ruppert2003semiparametric}.
We introduce now the conditions required for the definition and consistency
of our estimator.

\begin{description} 
\item [{Condition~C1}] \textbf{(a)} $\Theta$ is a compact set of $\mathbb{R}^{p}$ and $\theta^{\text{*}}$ is an interior point of $\text{\ensuremath{\Theta}}$, $\mathcal{X}$ is an open subset of $\mathbb{R}^{2}$ ; \textbf{(b)} $(t,x)\mapsto f(t,x,\theta^{*})$ is $L^{2}$-Lipschitz and $L^{2}$-Caratheodory (see \emph{Supplementary Material I}, section 1).

\item [{Condition~C2}] \textbf{(a)} $(Y_{i},t_{i})$ are i.i.d. with variance $V(Y\vert T=t)=\Sigma_{\epsilon}=\sigma^{2}I_{2}$ ; \textbf{(b)} For every $K$, there is a nonsingular constant matrix B such that for $P^{K}=B_{p}^{K}(t)$; \textbf{\emph{(i)}} the smallest eigenvalue of $E\left[P^{K}(T)P^{K}(T)^{\top}\right]$ is bounded away from zero uniformly in $K$ and \textbf{\emph{(ii)}} there is a sequence of constants $\zeta_{0}(K)$ satisfying $\sup_{t}\left|P^{K}(t)\right|\leq\zeta_{0}(K)$ and $K=K(n)$ such that $\zeta_{0}(K)^{2}K/n\longrightarrow0$ as $n\longrightarrow\infty$ ; \textbf{(c)} There are $\alpha,\mathbf{c}_{1,K},\mathbf{c}_{2,K}$ such that $\left\Vert \phi_{j}^{*}-p^{K}\mathbf{c}_{j,K}\right\Vert _{\infty}=\sup_{\left[0,1\right]}\left|\phi_{j}^{*}(t)-p^{K}(t)^{\top}\mathbf{c}_{j,K}\right|=O(K^{-\alpha})$. 

\item [{Condition~C3}] There exists $D>0$, such that the $D$-neighborhood of the solution range $\mathcal{D}=\{ x\in\mathbb{R}^{2} \vert $ $ \exists t\in [0,1], \vert x-\phi^{*}(t)\vert<D \} $ is included in $\mathcal{X}$ and $f$ is $C^{2}$ in $(x,\theta)$ on $\mathcal{D}\times\Theta$ for $t$ in $\left[0,1\right]$ a.e. Moreover, the derivatives of $f$ w.r.t $x$ and $\theta$ (with obvious notations) $f_{x}$, $f_{\theta}$, $f_{xx}$, $f_{x\theta}$ and $f_{\theta\theta}$ are $L^{2}$ uniformly bounded on $\mathcal{D}\times\Theta$ by $L^{2}$ functions $\bar{h}_{x}$, $\bar{h}_{\theta}$, $\bar{h}_{x\theta}$,$\bar{h}_{xx}$ and $\bar{h}_{\theta\theta}$ (respectively). 

\item [{Condition~C4}] Let $\left(\varphi_{\ell}\right)_{\ell\geq1}$ be an orthonormal sequence of $C^1$ functions in $H_0^1$. 

\item [{Condition~C5}] $\theta^{*}$ is the unique global minimizer of $Q_{\mathcal{F}}^{*}$ and $\inf_{\left|\theta-\theta^{*}\right|>\epsilon}Q_{\mathcal{F}}^{*}(\theta)>0$. 

\item [{Condition~C6}] There exists $L_{0}$ such that for $L\geq L_{0}$, $\mathbf{J}_{\theta,L}\left(g,\theta\right)$ is full rank in a neighborhood of $(\phi^{*},\theta^{*})$. 

\end{description}

Condition \textbf{C1} gives the existence and uniqueness of a solution
$\phi^{*}$ in $H^1$ to the IVP for $\theta=\theta^{*}$ and $x(0)=\phi^{*}(0)$.
If $f$ is continuous in $t$ and $x$, then the derivative $\dot{\phi^{*}}(t)=f\left(t,\phi^{*}(t),\theta^{*}\right)$
can be defined on $\left]0,1\right[$ and is also continuous, see appendix A. More generally, \textbf{C1} does apply when there is a discontinuous input variable, such as in the Ricatti example described in section \ref{sec:RicattiEx}.     \\

Under condition\textbf{ C2} (satisfied among others by regression
splines with $\zeta_{0}(K)=\sqrt{K}$), it is known that the series
estimator $\hat{\phi}_{j}$ are consistent estimators of $\phi_{j}^{*}$
for usual norms, in particular $\left\Vert \hat{\phi}_{j}-\phi_{j}^{*}\right\Vert _{\infty}=O_{P}\left(\zeta_{0}(K)\left(\sqrt{\nicefrac{K}{n}}+K^{-\alpha}\right)\right)$
(theorem 1, \cite{Newey1997}). If $\phi^{*}$ is $C^{s}$ and we
use splines then $\alpha=s$ and $\left\Vert \hat{\phi}-\phi^{*}\right\Vert _{\infty}=O_{P}\left(\nicefrac{K}{\sqrt{n}}+K^{1/2-s}\right)$. \\

Condition \textbf{C3} is here to control the continuity and regularity of the function $\mathcal{E}$ involved in the inverse problem. 
Moreover, it provides uniform control needed for stochastic convergence. \\

Condition \textbf{C4} is a sufficient condition for deriving independent conditions $\mathcal{C}_\ell(\theta)$, and normalization is useful only to avoid giving implicitly more weight to a condition w.r.t. the other conditions. %A principled choice of weights and orthogonal functions is discussed in section \ref{sub:OptimalVariance}. \\

Condition \textbf{C5} means that $\theta^{*}$ is a global and isolated
minima of $Q_{\mathcal{F}}^{*}(\theta)$, which is standard in M-estimation \cite{Vaart1998},
but can be hard to check in practice. Indeed, the parametric identifiability of ODE models can be hard to show, even for small systems. No general and practical results do exist for assessing the identifiability of an ODE model \cite{MiaoXiaPerelsonWu-siam2011}: it is useful to discriminate between ODE identifiability, statistical identifiability and practical identifiability. The latter being the most useful but almost impossible to check a priori.  
The essential meaning of condition \textbf{C5} is  that the addition of more and more orthogonal conditions should lead to a perfect and univocal estimation of the true parameter. From our experience and by numerical computations, we can check that $Q_L^{*}(\theta)$ has a unique minima in $\theta^*$ in a region of interest, for $L$ big enough (usually $L\simeq 2\times p$). 
The natural criterion for estimating $\theta$ and for identifiability analysis is 
\[
Q^{*}(\theta)=\left\Vert \mathcal{E}\left(\phi^{*},\theta\right)-\mathcal{E}\left(\phi^{*},\theta^{*}\right)\right\Vert _{L^{2}}^{2}
\]
but $\left\Vert \boldsymbol{E}_{\mathcal{F}}^{\bot}\left(\phi^{*},\theta\right)\right\Vert _{L^{2}}^{2}$ is withdrawn and 
we use the quadratic form $Q_{\mathcal{F}}^{*}(\theta)$ in order
to avoid boundary effects. This is needed in order to get a parametric rate of convergence, as in the original two-step criterion (\ref{eq:ClassicalTS}).   As a consequence, we lose a piece of information
brought by the trajectory $t\mapsto\phi^{*}(t)$ and we have to be
sure that the parameter $\theta$ has a low influence on $\left\Vert \boldsymbol{E}_{\mathcal{F}}^{\bot}\left(\phi^{*},\theta\right)\right\Vert _{L^{2}}^{2}$.
A favorable case is that it is almost constant on $\Theta$, so that
$Q^{*}$ and $Q_{\mathcal{F}}^{*}$ are essentially the same functions,
with the same global minimum and the same discriminating power. 
In practice, we can check that  \textbf{C5}  is approximately satisfied by computing numerically the criterion $\boldsymbol{E}_{L'}\left(\phi(\cdot,\hat{\theta}_{n,L}),\theta \right)$, in a neighborhood of $\hat{\theta}_{n,L}$,  for $L'\geq L$. 

Finally, Condition \textbf{C6} is about the influence of the number of test functions used. We use only the first $L$ Fourier coefficients
of $\mathcal{E}(g,\theta)-\dot{g}$ to identify the parameter $\theta$,
but this might not be sufficient to discriminate between two parameters
$\theta$ and $\theta'$. In a way, we perform dimension reduction
but we need to be sure that we have an exact recovery when $L$ goes
to infinity: we expect that the global minimum $\theta_L^{*}$ of $\left|\mathbf{e}_{L}^{*}(\theta)\right|^{2}$
 is close to the global minimum $\theta{*}$ of $Q_{\mathcal{F}}^{*}(\theta)=\left\Vert \boldsymbol{E}_{\mathcal{F}}\left(\phi^{*},\theta\right)\right\Vert _{L^{2}}^{2}$ (found under condition \textbf{C5}). 
We introduce the Jacobian matrices $\mathbf{J}_{\theta,L}\left(g,\theta\right)$
in $\mathbb{R}^{L\times p}$ with entries $\int_{0}^{1}f_{\theta_{j}}(t,g(t),\theta)\varphi_{\ell}(t)dt$
and $ $$\mathbf{J}_{x,L}\left(g,\theta\right)$ in $\mathbb{R}^{L\times d}$
with entries $\int_{0}^{1}f_{x_{i}}(t,g(t),\theta)\varphi_{\ell}(t)dt$.
For this reason, we suppose that $\mathbf{J}_{\theta,L}^{*}$ is full
rank, so that $Q_{L}^{*}(\theta)$ is locally strictly convex, with
a unique local minimum $\theta_L^{*}$. \\
The Jacobian matrix introduced in condition \textbf{C6} is classical in sensitivity analysis (in  ODE models). Usually, the sensitivity matrix used is the Jacobian of the least squares criterion (similar to $\mathbf{J}_{\theta,L}\left(\phi(\cdot,\hat{\theta}),\theta\right)$); it enables to check a posteriori the identifiability of the parameter $\theta$. Conversely, local non-identifiable parameter (\emph{sloppy parameters}, \cite{Sethna2007}) can be detected in that case. 

\begin{thm}
\label{thm:Consistency}If conditions \textbf{C1} to \textbf{C6} are satisfied, then 
\[
\hat{\theta}_{n,L}-\theta_L^{*}=O_{P}(1)
\]
and the bias $\boldsymbol{B}_{L}=\theta_L^{*}-\theta^*$ tends to zero as $L\rightarrow\infty$. \\
In particular, if we use the sine basis and if $\mathcal{E}\left(\phi^{*},\theta\right)$ is in $H^{1}$ for all $\theta$, then $\boldsymbol{B}_{L}=o\left(\frac{1}{L}\right)$.
\end{thm}

\begin{remark}
The convergence rate of the bias $\boldsymbol{B}_L$ can be refined according to the test functions $\varphi_{\ell}$: if we use B-splines, the bias is controlled by the meshsize $\Delta=\max_{j>1} (\tau_j-\tau_{j-1})$  of the sequence of knots $\tau_j,\, j=1,\dots,L$ defining the spline spaces, see section 6 in \cite{Schumaker2007}. 
\end{remark}
\begin{remark}
In practice, we have $\boldsymbol{B}_L=0$ for medium-size $L$, around $2\times d \times p$. 
\end{remark}

\section{Asymptotics\label{sec:Asymptotics}}

We give a precise description of the asymptotics of $\hat{\theta}_{n,L}$
(rate, variance and normality), by exploiting the well-known properties
of series estimators. We consider the linear case, then we extend
the obtained results to general nonlinear ODEs. We show in a preliminary
step that the asymptotics of $\hat{\theta}_{n,L}-\theta_{L}^{*}$
are directly related to the behavior of $ $$\mathbf{e}_{L}(\hat{\phi},\theta^{*})$,
which is a classical feature of Moment Estimators.

\subsection{Asymptotic representation for $\hat{\theta}_{n}-\theta_L^{*}$ \label{sub:Asymptotic-representation}}

From the definition (\ref{eq:OrthogonalEstimator}) of $\hat{\theta}_{n,L}$
and differentiability of $f$, the first order optimality condition
is 
\begin{equation}
\mathbf{J}_{\theta,L}\left(\hat{\phi},\hat{\theta}_{n,L}\right)^{\top}\mathbf{e}_{L}\left(\hat{\phi},\hat{\theta}_{n,L}\right)=0\label{eq:FirstOrderCondition}
\end{equation}
from which we derive an asymptotic representation for $\hat{\theta}_{n,L}$,
by linearizing $\mathbf{e}_{L}\left(\hat{\phi},\hat{\theta}_{n,L}\right)$
around $\theta_{L}^{*}$. We need to introduce the matrix-valued function defined
on $\mathcal{D}\times\theta$ such that $\boldsymbol{M}_{L}(g,\theta)=\left[\mathbf{J}_{\theta,L}\left(g,\theta\right)^{\top}\mathbf{J}_{\theta,L}\left(g,\theta\right)\right]^{-1}$ $\mathbf{J}_{\theta,L}\left(g,\theta\right)^{\top}$,
and proposition \ref{pro:LinkParameter2Conditions} shows that $ $$\boldsymbol{M}_{L}(\hat{\phi},\hat{\theta}_{n,L})$
is also a consistent estimator of $\boldsymbol{M}_{L}^{*}$. 

\begin{prop}
\label{pro:LinkParameter2Conditions}If conditions \textbf{C1}-\textbf{C6}
are satisfied, then
\begin{equation}
\left[\mathbf{J}_{\theta,L}\left(\hat{\phi},\hat{\theta}_{n,L}\right)^{\top}\widetilde{\mathbf{J}}_{L}\right]^{-1}\mathbf{J}_{\theta,L}\left(\hat{\phi},\hat{\theta}_{n,L}\right)^{\top}\stackrel{P}{\longrightarrow}\boldsymbol{M}_{L}^{*}=\left[\mathbf{J}_{\theta,L}^{*\top}\mathbf{J}_{\theta,L}^{*}\right]^{-1}\mathbf{J}_{\theta,L}^{*\top}\label{eq:DefinitionmatM}
\end{equation}
where the matrix $\widetilde{\mathbf{J}}_{L}$ is the Jacobian $\mathbf{J}_{\theta,L}$ evaluated at a point $\widetilde{\theta}$ between $\theta^*$ and $\hat{\theta}_{n,L}$. 
Moreover,  we hav\textup{e
\begin{equation}
\hat{\theta}_{n,L}-\theta_{L}^{*}=-\boldsymbol{M}_{L}^{*}\mathbf{e}_{L}(\hat{\phi},\theta_L^{*})+o_{P}(1).\label{eq:AsympRepEstimator}
\end{equation}
}
\end{prop}

\subsection{Linear differential equations\label{sub:Linear-differential-equations}}

We consider the parametrized linear ODE defined as

\begin{equation}
\begin{cases}
\dot{x}_{1} & =a(t,\theta_{1})x_{1}+b(t,\theta_{1})x_{2}\\
\dot{x}_{2} & =c(t,\theta_{2})x_{1}+d(t,\theta_{2})x_{2}
\end{cases}\label{eq:OdeLinear2D}
\end{equation}
%Since the ODE is linear, conditions \textbf{IVP(a)} and \textbf{IVP(b)}
%are satisfied as soon as the functions 
where $a(\cdot,\theta)$, $b(\cdot,\theta)$, $c(\cdot,\theta)$, $d(\cdot,\theta)$
are in $L^{2}$. 
We focus only on the estimation of the parameter $\theta=\theta_{1}$ involved in
the first equation $\dot{x}_{1}=a(t,\theta)x_{1}+b(t,\theta)x_{2}$
and we suppose that we have two series estimators $\hat{\phi}_{1}=\boldsymbol{p}_{K}^{\top}\hat{\mathbf{c}}_{1}$
and $\hat{\phi}_{2}=\boldsymbol{p}_{K}^{\top}\hat{\mathbf{c}}_{2}$
satisfying condition \textbf{C2}. The orthogonal conditions are simple linear functionals of the estimators 
$e_{\ell}(\hat{\phi},\theta) = \left\langle \hat{\phi}_{1},\dot{\varphi_{\ell}}+a(\cdot,\theta)\varphi_{\ell}\right\rangle +\left\langle \hat{\phi}_{2},b(\cdot,\theta)\varphi_{\ell}\right\rangle$.
Hence the asymptotic behavior of the empirical orthogonal conditions
relies on the plug-in properties of $\hat{\phi}_{1}$ and $\hat{\phi}_{2}$
into the linear forms $T_{\rho}:x\mapsto\int_{0}^{1}\rho(t)x(t)dt$
where $\rho$ is a smooth function. Moreover, the linearity of series
estimator makes the orthogonal conditions $\mathbf{e}_{L}(\hat{\phi},\theta)$ easy to compute as \begin{equation}
\mathbf{e}_{L}(\hat{\phi},\theta)=\mathbf{A}(\theta)\hat{\mathbf{c}}_{1}+\mathbf{B}(\theta)\hat{\mathbf{c}}_{2}\label{eq:LinearOrthoCOnditions}
\end{equation}
where $\mathbf{A}(\theta)$ and $\mathbf{B}(\theta)$ are matrices
in $\mathbb{R}^{L\times K}$ with entries $A{}_{\ell,k}(\theta) =  \int_{0}^{1}\left(a(t,\theta)\varphi_{\ell}(t)+\dot{\varphi}_{\ell}(t)\right)p_{kK}(t)dt$ and $B_{\ell,k}(\theta)=\int_{0}^{1}\left(b(t,\theta)\varphi_{\ell}(t)\right)p_{kK}(t)dt$.
% \[
% \begin{array}{lll}
% A{}_{\ell,k}(\theta) & = & \int_{0}^{1}\left(a(t,\theta)\varphi_{\ell}(t)+\dot{\varphi}_{\ell}(t)\right)p_{kK}(t)dt\\
% B_{\ell,k}(\theta) & = & \int_{0}^{1}\left(b(t,\theta)\varphi_{\ell}(t)\right)p_{kK}(t)dt
%\end{array}
%\]
The gradient of $\mathbf{e}_{L}(\hat{\phi},\theta)$ is $\mathbf{J}_{\theta,L}\left(\hat{\phi},\theta\right)=\partial_{\theta}\mathbf{A}(\theta)\hat{\mathbf{c}}_{1}+\partial_{\theta}\mathbf{B}(\theta)\hat{\mathbf{c}}_{2}$
where $\partial_{\theta}\mathbf{A}(\theta)$ and $\partial_{\theta}\mathbf{B}(\theta)$
are straightforwardly computed by permuting differentiation and integration. 
%\[
%\begin{array}{lll}
%\partial_{\theta}\mathbf{A}(\theta)_{kl} & = & \int_{0}^{1}\partial_{\theta}a(t,\theta)\varphi_{\ell}(t)p_{kK}(t)dt\\
%\\
%\partial_{\theta}\mathbf{B}(\theta)_{kl} & = & \int_{0}^{1}\partial_{\theta}b(t,\theta)\varphi_{\ell}(t)p_{kK}(t)dt
%\end{array}.
%\]
Although $\mathbf{e}_{L}(\hat{\phi},\theta)$ depends linearly on
the observations, we have to take care of the asymptotics as we are
in a nonparametric framework and $K$ grows with $n$. The behavior
of linear functionals $T_{\rho}(\hat{\phi})$ for several nonparametric
estimators (kernel regression, series estimators, orthogonal series)
is well known \cite{Andrews1991,BickelRitov2003,Goldstein92,Newey1997},
and in generality it can be shown that such linear forms can be estimated
with the classical root-$n$ rate and that they are asymptotically
normal under quite general conditions. In the particular case of series
estimators, we rely on theorem 3 of \cite{Newey1997} that ensures
the root-$n$ consistency and the asymptotic normality of the plugged-in
estimators $T_{\rho}(\hat{\phi}_{j}),\, j=1,2$ under almost minimal
conditions. We will give in the next section the precise assumptions
required for root-$n$ consistency of linear and nonlinear functional
of the series estimator. Moreover, the variance of $\hat{\theta}_{n,L}$ has a remarkable expression
\begin{equation}
V_{e,L}(\theta)=V\left(\mathbf{e}_{L}(\hat{\phi},\theta)\right)=\mathbf{A}(\theta)V\left(\hat{\mathbf{c}}_{1}\right)\mathbf{A}(\theta)^{\top}+\mathbf{B}(\theta)V(\hat{\mathbf{c}}_{2})\mathbf{B}(\theta)^{\top}.\label{eq:VarianceLinearCond}
\end{equation}
We remark that there is no covariance term between $\hat{\mathbf{c}}_{1}$
and $\hat{\mathbf{c}}_{2}$ since we assume that $V\left(Y\vert T=t\right)$
is diagonal (assumption \textbf{C2}), but in all generality, we should add
$2\mathbf{A}(\theta)\mbox{cov}(\hat{\mathbf{c}}_{1},\hat{\mathbf{c}}_{2})\mathbf{B}(\theta)^{\top}$.
We can use the classical estimates of the variance of $\hat{\mathbf{c}}_{1}$
and $\hat{\mathbf{c}}_{2}$ to compute an estimate of $V_{e,L}(\theta)$

\begin{equation}
\widehat{V_{e,L}(\theta)}=\mathbf{A}(\theta)\widehat{V\left(\hat{\mathbf{c}}_{1}\right)}\mathbf{A}(\theta)^{\top}+\mathbf{B}(\theta)\widehat{V(\hat{\mathbf{c}}_{2})}\mathbf{B}(\theta)^{\top}\label{eq:EmpVarianceLinearCond}
\end{equation}
Thanks to proposition \ref{pro:LinkParameter2Conditions}, we can
estimate the asymptotic variance of the estimator $\hat{\theta}_{n,L}$
with the consistent estimator $\hat{\boldsymbol{M}}_{L}=\boldsymbol{M}_{L}(\hat{\phi},\hat{\theta}_{n,L})$
and we estimate $V\left(\hat{\theta}_{n,L}\right)$ by $\widehat{V\left(\hat{\theta}_{n,L}\right)}=\hat{\boldsymbol{M}}_{L}\widehat{V\left(\mathbf{e}_{L}(\hat{\phi},\hat{\theta}_{n,L})\right)}\hat{\boldsymbol{M}}_{L}^{\top}$.
From the asymptotic normality of the plug-in estimate, we can derive
confidence balls with level $1-\alpha$. For instance,
for each parameter $\theta_{i}$, $i=1,\dots,p$: 
\[
IC(\theta_{i};1-\alpha)=\left[\left(\hat{\theta}_{n,L}\right)_{i}\pm q_{1-\frac{\alpha}{2}}\widehat{V\left(\hat{\theta}_{n,L}\right)}_{ii}^{1/2}\right]
\]
where $q_{1-\alpha/2}$ is the quantile of order $1-\frac{\alpha}{2}$
of a standard Gaussian distribution. Nevertheless, we recall that
these confidence intervals might be affected by the bias of $\hat{\theta}_{n,L}$
depending on $L$.
\begin{description}
\item [{}]~
\end{description}

\subsection{Nonlinear differential equations\label{sub:Nonlinear-differential-equations}}

We give here general results for the asymptotics of $ $$e_{\ell}(\hat{\phi},\theta)$
when the functional is linear or not in $\hat{\phi}$. In \cite{Newey1997},
the root-$n$ consistency and asymptotic normality is obtained if
the functional $g\mapsto e_{\ell}(g,\theta)$ has a continuous Fr\'echet
derivative $De_{\ell}(g,\theta)$ with respect to the norm $\left\Vert \cdot\right\Vert _{\infty}$.
If $x\mapsto f\left(t,x,\theta\right)$ is twice continuously differentiable
for $t\in\left[0,1\right]$ a.e. in $x$ and $\theta$ in $ $$\Theta$, then
we can compute easily its Fr\'echet derivative for $g\in H^{1}$ in
the uniform ball $\left\Vert g-\phi^{*}\right\Vert _{\infty}\leq D$
. For all $h\in H^{1}$ such $\left\Vert g+h-\phi^{*}\right\Vert _{\infty}\leq D$,
we have \begin{equation}
e_{\ell}(g+h,\theta)-e_{\ell}(g,\theta) = \left\langle f_{x}\left(\cdot,g,\theta\right)h,\varphi_{\ell}\right\rangle +\left\langle h,\dot{\varphi}\right\rangle +\left\langle h^{\top}f_{xx}\left(\cdot,\tilde{g},\theta\right)h,\varphi_{\ell}\right\rangle \label{eq: FrechetDev} 
\end{equation}
by a Taylor expansion around $g$. As in the linear case, we introduce
the tangent linear operator $\mathcal{A}_{g}(\theta)\,:\, u\mapsto\dot{u}-a_{g}(t,\theta)u$
with $a_{g}(t,\theta)=f_{x_{1}}(t,g(t),\theta)$ and the function $b_{g}(t,\theta)=f_{x_{2}}(t,g(t),\theta)$.
% the adjoint operator $\mathcal{A}_{g}^{*}(\theta)\,:\, u\mapsto\dot{u}+a_{g}(t,\theta)u$
For all $\theta$, the Fr\'echet derivative of $e_{\ell}(g,\theta)$ (w.r.t
to the uniform norm) is the linear operator $h=(h_{1},h_{2})\text{\ensuremath{\mapsto}}De_{\ell}(g,\theta).h=$  $\left\langle h_{1}, \dot{\varphi_{\ell}}+a_{g}(t,\theta)\varphi_{\ell}\right\rangle +\left\langle h_{2},b_{g}(\cdot,\theta)\varphi_{\ell}\right\rangle $
and satisfies for all $\theta\in\Theta$
\[
\left|e_{\ell}(g+h,\theta)-e_{\ell}(g,\theta)-De_{\ell}(g,\theta).h\right|\leq C\left\Vert h\right\Vert _{\infty}^{2}
\]
because $f_{xx}$ is uniformly dominated on $\mathcal{D}\times\Theta$.
Moreover, for all $\epsilon$ (with $0<\epsilon<D$), for all $g,g'$
such that $\left\Vert g-\phi^{*}\right\Vert _{\infty},\left\Vert g'-\phi^{*}\right\Vert _{\infty}\leq\epsilon$,
we have
\begin{eqnarray*}
\left|De_{\ell}(g,\theta).h-De_{\ell}(g',\theta).h\right| & \leq & \int_{0}^{1}h(t)^{\top}f_{xx}\left(t,\tilde{g}(t),\theta\right)\left(g(t)-g'(t)\right)\varphi_{\ell}(t)dt\\
 & \leq & C\left\Vert h\right\Vert _{\infty}\left\Vert g-g'\right\Vert _{\infty}
\end{eqnarray*}
with $C$, a constant independent of $\theta$, $\epsilon$ and $g,g'$
(because $f_{xx}$ is uniformly dominated). 

As in the linear case, we need to evaluate $De_{\ell}(g,\theta)$
on the basis $\boldsymbol{p}^{K}$. We denote $\mathbf{A}(g,\theta)$
and $\mathbf{B}(g,\theta)$ the matrices in $\mathbb{R}^{L\times K}$
with entries $\int_{0}^{1}a_{g}(t,\theta)\varphi_{\ell}(t)p_{kK}dt$
and $\int_{0}^{1}b_{g}(t,\theta)\varphi_{\ell}(t)p_{kK}dt$ (respectively)
and we have the approximation 
\begin{equation}
\mathbf{e}_{L}(\hat{\phi},\theta)=\mathbf{e}_{L}(\phi^{*},\theta)+\mathbf{A}(\phi^{*},\theta)\hat{\mathbf{c}}_{1}+\mathbf{B}(\phi^{*},\theta)\hat{\mathbf{c}}_{2}+O\left(\left\Vert h\right\Vert _{\infty}^{2}\right).\label{eq:NonLinearOrthoConditions}
\end{equation}
We can derive the asymptotic variance of $\mathbf{e}_{L}(\hat{\phi},\theta)$
from (\ref{eq:NonLinearOrthoConditions}) 
\begin{eqnarray}
V_{e,L}(\theta) & = & \mathbf{A}(\phi^{*},\theta)V\left(\hat{\mathbf{c}}_{1}\right)\mathbf{A}(\phi^{*},\theta)^{\top}+\mathbf{B}(\phi^{*},\theta)V(\hat{\mathbf{c}}_{2})\mathbf{B}(\phi^{*},\theta)^{\top}\label{eq:AsymptotVarianceConditions}
\end{eqnarray}
and we can get an estimate $\widehat{V_{e,L}(\theta)}$ from the data
as in the linear case.\\ 

In order to assess the previous discussion and for deriving the root-$n$ rate of our estimator, we introduce the following two conditions: 
\begin{description}
\item [{Condition~C7}] \textbf{(a)} The times $T_{1},\dots,T_{n}$ have a density $\pi$ w.r.t. Lebesgue measure such $0<c<\pi<C<\infty$; \textbf{(b)} $E\left[\epsilon^{4}\right]<\infty$.  

\item [{Condition~C8}] For $\ell=1,\dots,L$, $\theta \in \Theta$, there exists $\tilde{\beta}_{K_\ell}$ in $\mathbb{R}^{K_\ell}$ with $ 
\Vert \frac{f_{x}\left(\cdot,\phi^{*},\theta\right)\varphi_{\ell}+\dot{\varphi}_{\ell}}{\pi}-\tilde{\beta}_{K_\ell}^{\top}p^{K_\ell}\Vert_{L^2}\longrightarrow 0$. 
\end{description}
Conditions \textbf{C7} and \textbf{C8}  are similar to the assumptions given in \cite{Newey1997}. Condition \textbf{C8} is here to ensure that the Fr\'echet derivative $De_{\ell}(\phi^*,\theta)$ that drives the asymptotic rate of $e_{\ell}(g,\theta)$ (see equation \ref{eq: FrechetDev}) can be well approximated in the basis $\boldsymbol{p}^K$ as the nonparametric proxy. Then the linearized nonlinear functional of the nonparametric estimator is well approximated by a linear combination of the regression coefficients. When we use B-splines with uniform knot sequence, condition \textbf{C8} can be replaced by the simpler condition \textbf{C9}:

\begin{description}
\item [{Condition~~C9}] \textbf{(a)} The series estimator is a regression
spline with a uniform knot sequence $(\tau_{1,K},\dots,\tau_{N_{K},K})$
defining the spline basis $\boldsymbol{p}^{K}$ satisfies $\max_{i}\vert\tau_{i+1,K}-\tau_{i,K}\vert\longrightarrow0$
as $K\longrightarrow \infty$ ; \textbf{(b)} For all $\theta\in\Theta$,
for $\ell=1\dots L$, $v_{\ell}\,:\, t\mapsto\frac{f_{x}\left(t,\phi^{*}(t),\theta\right)\varphi_{\ell}(t)+\dot{\varphi}_{\ell}(t)}{\pi(t)}$
is $C^{1}$. \end{description}

\begin{thm}\label{thm:Rootn} 
If either the following conditions are satisfied:
\begin{description}
\item[$\boldsymbol{p}^K$ is a general series estimators] Under conditions \textbf{C1-C8} and if $f$ is a linear vector field or, $f$ is a nonlinear vector field
and $K$ is chosen such that $\frac{\zeta_{0}(K)^{4}K^{2}}{n}\longrightarrow 0$ 
\item[$\boldsymbol{p}^K$ is a uniform knot splines] Under conditions \textbf{C1-C2(a)},\textbf{C3-C7},\textbf{C9} and if $f$ is a linear vector field and $\frac{K^2}{n}\longrightarrow 0$, or $f$ is a nonlinear vector field and $\frac{K^{4}}{n}\longrightarrow 0$
\end{description}
Then $\hat{\theta}_{n,L}$ is such that 
\begin{equation} \sqrt{n}\left(\hat{\theta}_{n,L}-\theta_{L}^{*}\right)\rightsquigarrow N(0,\mathbf{V}_{L}^{*})\label{eq:AsymptNormalityParameter} \end{equation}
with 
\begin{equation} \mathbf{V}_{L}^{*}=\mathbf{M}_{L}^{*}\mathbf{V}_{e,L}^{*}\mathbf{M}_{L}^{*\top}.\label{eq:VarianceOrthogonalConditions} \end{equation}
where $\mathbf{V}_{e,L}^{*}=V_{e,L}\left(\theta_L^{*}\right)$. The asymptotic variance can be estimated by
$\hat{\boldsymbol{M}}_{L}\widehat{V_{e,L}(\hat{\theta}_{n,L})}\hat{\boldsymbol{M}}_{L}^{\top}\stackrel{P}{\longrightarrow}\mathbf{V}_{L}^{*}$. 
In particular, if we use regression splines and $t \mapsto f(t,\phi^{*}(t),\theta)$ is $C^{s}$ on $\left[0,1\right]$ with $s\geq3$, then (\ref{eq:AsymptNormalityParameter}) holds with $K$ such that $\sqrt{n}K^{-s}\rightarrow0$ and $n^{-1}K^{4}\rightarrow0$.  \\Moreover, if $L=L(n)\longrightarrow \infty, n\longrightarrow \infty$  is chosen such that  the bias $\boldsymbol{B}_{L(n)}=O(n^{-1/2})$, then
we have 
\begin{equation}
\hat{\theta}_{n,L(n)}-\theta^{*}=O_{P}(n^{-1/2}).\label{eq:rootn-withoutbias}
\end{equation}
In particular, this is the case when the test  functions $\varphi_{\ell}$ are the sine basis, and  $L(n)=O(n^{\alpha})$ with $\alpha>1/2$. 
\end{thm}

This theorem is a direct application of theorem 3 in
\cite{Newey1997} that claims the root-$n$ consistency and asymptotic
normality of general (nonlinear) plug-in estimators. The main steps of the proof are given in  \emph{Supplementary Material I}.

\section{Experiments}

\subsection{Description of the setting}

We compare the NLS estimator $\hat{\theta}^{NLS}$, the Two-Step Estimator
(TS) $\hat{\theta}^{TS}$ and the OC estimator $\hat{\theta}^{OC}$
for varying sample sizes ($n=400,\,200,\,50$) and varying noise levels
(high and small). We consider 3 different ODEs with different mathematical
structure: the $\alpha$-pinene ODE (linear in state and in parameter), the Ricatti ODE (nonlinear in state, linear in parameter) and the FitzHugh-Nagumo ODE (nonlinear in state and in parameter).
These three models give a gross picture of the robustness, consistency
and efficiency of the different estimators. This can be critical as
the asymptotics are obtained by linearization and that the quality
of this approximation (in particular for the computation of the covariance
matrix) depends on the discrepancy with respect to linearity. 

In the simulations, the noise is homoscedastic and Gaussian, so that
the NLS are asymptotically efficient. Hence, the settings $n=200$
or $n=400$ indicates the efficiency loss of the Gradient Matching
estimators whereas the small size setting ($n=50$) gives some information
on the small sample case, where the asymptotic approximations cannot
be assessed. 

As the standard reference method, the Sum of Squared Errors (SSE)
is minimized by a Levenberg-Marquardt algorithm using 20 starting points
centered around the true parameter value $\theta^{*}$, and we retain
the best minimum. The solution of the ODE is computed by a Runge-Kutta
algorithm of order 4, implemented in the Matlab function \emph{ode45.}
Hence, we expect that we obtain the true NLS estimator, and that the
estimated variance is the true best one. 

The Gradient Matching estimators (TS and OC) use the same regression
spline, decomposed on a B-spline basis with a uniform knots sequence
$\xi_{k},k=1,\dots,K$. For each dataset (and each dimension), the
number of knots is selected by minimizing the GCV criterion, \cite{ruppert2003semiparametric}.
For the plain TS estimator, we use a piecewise affine weight function
with $w(0)=w(1)=0$, as in \cite{Brunel2008}.

The Orthogonal Conditions are defined with the sine basis or B-Splines
basis.  
%$\widehat{\theta}^{OC}$ is computed with a weight matrix $W=I_{L}$. The OC estimator computed with an estimated optimal is considered in \emph{Supplementary Material II}.
%and $\widehat{\theta}_{opt}^{OC}$ corresponds to the optimal weighting 
%matrix $W^{opt}$ given in proposition 4.2 (computed by IRWOC). For each estimate,
We have to face with the practical problem of finding
the best number of conditions $L$, that depends on the model and
on $\hat{\phi}$. In each setting, we have fixed a minimum and a maximum
number of conditions $\boldsymbol{L}_{min}$ and $\boldsymbol{L}_{max}\leq2\times d\times p$
and we select the OC estimator $\hat{\theta}_{n,L}$ that gives the
smallest prediction error (i.e that minimizes the SSE): 
\[
\hat{\theta}^{OC}=\arg\min_{\boldsymbol{L}_{min}\leq L\leq\boldsymbol{L}_{max}}\sum_{i=1}^{n}\left\Vert y_{i}-\phi(t_{i},\hat{\phi}_{0},\hat{\theta}_{n,L})\right\Vert ^{2}
\]
where $\hat{\phi}_{0}=\hat{\phi}(0)$ is the nonparametric estimate
of the initial condition. 
%The selection of $L$ is done separately
%for $\widehat{\theta}^{OC}$ and $\widehat{\theta}_{opt}^{OC}$ which
%induces that the results reported for the two OC estimators do not
%have necessarily the same number of conditions. This is partly due
%to the fact that the convergence of IRWOC requires to reduce the number
%of conditions, in order to avoid the singularity of the covariance
%matrix or because the linearization of the OC criterion is not enough
%accurate. 

We use Monte Carlo simulations, based on $N_{MC}=500$ independent
draws for comparing the estimators. We compute their Mean Squared
Errors $\left\Vert \hat{\theta}-\theta^{*}\right\Vert ^{2}$. 
% and their
%Absolute Relative Errors (ARE)
%\[
%ARE=\frac{1}{N_{MC}}\sum_{i=1}^{N_{MC}}\frac{\left|\theta^{*}-\widehat{\theta}_{i}\right|}{\left|\theta^{*}\right|}.
%\]
The accuracy of the estimator is roughly estimated by the trace of
the covariance matrices of the estimators, denoted $Tr\left(V(\hat{\theta})\right)$.
Moreover, the reliability of the estimates (and asymptotic approximation)
is evaluated with the coverage probabilities of the $95\%$ confidence
ellipse (except in the case of TS because there is no closed-form
for asymptotic variance). For the NLS, the asymptotic variance is
computed via the Matlab function \emph{nlinfit. }
A more detailed analysis of the experiments (including coverage probabilities of confidence sets) are given in \emph{Supplementary Materials II: Experiments, Tables and Figures}.  

\subsection{$\alpha$-pinene \label{sub:Linear-ODEs}}

A linear ODE with constant coefficients is written $\dot{x}=\boldsymbol{A}x$,
where $\boldsymbol{A}^{\top}=\left(A_{1}\vert\dots\vert A_{d}\right)$.
For $i=1,\dots,d$, the weak formulation gives the identity $\boldsymbol{Y}_{i}^{\varphi}=\boldsymbol{X}^{\varphi}A_{i}$
to be satisfied, where $\boldsymbol{X}^{\varphi}$ is a $d\times L$
matrix with entries $\left\langle x_{k},\varphi_{\ell}\right\rangle $
and$\boldsymbol{Y}_{i}^{\varphi}$ is a vector in $\mathbb{R}^{L}$
with entries equal to $-\left\langle x_{i},\dot{\varphi}_{\ell}\right\rangle $.
For illustration, we consider the \emph{$\alpha$-pinene ODE} used
in \cite{Moles2003} for the comparison of several global optimization
algorithms:

\begin{equation}
\left\{ \begin{array}{lll}
\dot{x}_{1} & = & -(\theta_{1}+\theta_{2})x_{1}\\
\dot{x}_{2} & = & \theta_{1}x_{1}\\
\dot{x}_{3} & = & \theta_{2}x_{1}-(\theta_{3}+\theta_{4})x_{3}+\theta_{5}x_{5}\\
\dot{x}_{4} & = & \theta_{3}x_{3}\\
\dot{x}_{5} & = & \theta_{4}x_{3}-\theta_{5}x_{5}
\end{array}\right.\label{eq:ode_alphapinen}
\end{equation}
The true parameter to be estimated from a completely observed trajectory
on $\left[0,100\right]$ is $\theta^{*}=\left(\theta_{1},\theta_{2},\theta_{3},\theta_{4},\theta_{5}\right)^{\top}$.
As this ODE is linear and time-invariant, we have a closed-form for
the solution $\phi^{*}(t,\theta,\phi_{0})=e^{tA}\phi_{0}$ that can
be directly used for the computation of the NLS estimator. 

The test functions used for the OC estimators are B-Splines (with
uniform knots sequence) $\varphi_{\ell},\ell=1,\dots,L$ with compact
support included in $\left]0,20\right[$. We consider a varying number
of conditions $L$, i.e $2\leq L\leq15$. Finally, we have two settings
for the estimation of $\theta$: when the initial condition $\phi_{0}$
is known (and equal to $(100,0,0,0,0)^{\top}$ as in \cite{Rodriguez2006}), and when $\phi_{0}$ is unknown and needs to be estimated
(for NLS).

\subsubsection{Known initial condition}

For the OC and TS estimator, we constrain the spline estimator $\hat{\phi}$
to satisfy the condition $\hat{\phi}(0)=\phi_{0}$ (by adding a linear
constraint to the classical least-squares minimization). Moreover,
following section 2.3, we integrate the knowledge of the initial condition
by adding a test function $\varphi_{0}$ which is a B-spline with
$\varphi_{0}(0)\neq0$. Hence, we define 2 differents OC estimators,
respectively, $\hat{\theta}^{OC,0}$ and $\hat{\theta}^{OC}$ that
uses or not (resp.) the knowledge of the initial condition. 

\begin{center}
\begin{table}[h]
\begin{centering}
\begin{tabular}{|c|c|c|c|c|c|c|c|c|}
\hline 
$\times10^{-2}$ & \multicolumn{4}{c|}{$MSE$} &  & \multicolumn{3}{c|}{$Tr\left(V(\hat{\theta})\right)$}\tabularnewline
\hline 
\hline 
$(n,\sigma)$ & TS & OC & OC,0 & NLS &  & OC & OC,0 & NLS\tabularnewline
\hline 
$(400,3)$ & 0.72 & 0.05 & 0.04 & 0.02 &  & 0.04 & 0.04 & 0.02\tabularnewline
\hline 
$(400,8)$ & 2.28 & 0.22 & 0.25 & 0.10 &  & 0.95 & 1.20 & 0.12\tabularnewline
\hline 
$(200,3)$ & 1.19 & 0.27 & 0.30 & 0.03 &  & 0.09 & 0.13 & 0.03\tabularnewline
\hline 
$(200,8)$ & 2.95 & 0.44 & 0.37 & 0.18 &  & 2.66 & 2.68 & 0.27\tabularnewline
\hline 
$(50,3)$ & 2.39 & 0.27 & 0.26 & 0.16 &  & 1.37 & 1.58 & 0.16\tabularnewline
\hline 
$(50,8)$ & 4.54 & 1.03 & 0.93 & 0.68 &  & 7.96 & 7.27 & 1.68\tabularnewline
\hline 
\end{tabular}
\par\end{centering}

\caption{MSE, Asymptotic Variance for \foreignlanguage{american}{$\alpha$-pinene
model with known Initial Condition\label{tab:mse_are_a_pinene_known_ci}}}
\end{table}

\par\end{center}

\subsubsection{Unknown initial condition}

In this case, the NLS needs to estimate the initial condition as well,
whereas it is not needed for Gradient Matching estimators and we have
the same estimates (for $\hat{\theta}^{TS}$ and $\hat{\theta}^{OC}$)
as in the previous section. In this setting, we consider another OC
estimator that uses information about the other boundary $T=100$.
Indeed, we know that the $\alpha-$pinene network converges to a stationary
point, that is almost reached at time $T=100$. Hence the boundary
condition $\dot{\phi}^{*}(100)=0$ can be used for estimation (section
2.3): if $\varphi_{1}$ is a test function with $\varphi_{1}(100)\neq0$,
we have $A^{2}<\phi^{*},\varphi_{1}>+A<\phi^{*},\dot{\varphi}_{1}>=0$.
This gives an additional condition to be satisfied for the OC estimator,
which is denoted as $\hat{\theta}^{OC,1}$, see section 2.3. 

\begin{center}
\begin{table}[h]
\begin{centering}
\begin{tabular}{|c|c|c|c|c|c|c|c|c|}
\hline 
$\times10^{-2}$ & \multicolumn{4}{c|}{$MSE$} &  & \multicolumn{3}{c|}{$Tr\left(V(\hat{\theta})\right)$}\tabularnewline
\hline 
\hline 
$(n,\sigma)$ & TS & OC & OC,1 & NLS &  & OC & OC,1 & NLS\tabularnewline
\hline 
$(400,3)$ & 0.25 & 0.11 & 0.11 & 0.07 &  & 0.10 & 0.10 & 0.06\tabularnewline
\hline 
$(400,8)$ & 1.07 & 0.85 & 0.56 & 0.50 &  & 1.06 & 0.82 & 0.61\tabularnewline
\hline 
$(200,3)$ & 0.6 & 0.37 & 0.23 & 0.14 &  & 0.25 & 0.20 & 0.14\tabularnewline
\hline 
$(200,8)$ & 1.64 & 1.42 & 0.83 & 1.34 &  & 2.36 & 1.64 & 1.54\tabularnewline
\hline 
$(50,3)$ & 1.33 & 1.31 & 0.80 & 0.69 &  & 1.63 & 1.02 & 0.76\tabularnewline
\hline 
$(50,8)$ & 3.64 & 2.11 & 1.79 & 1.96 &  & 5.34 & 2.20 & 4.38\tabularnewline
\hline 
\end{tabular}
\par\end{centering}

\caption{MSE, Asymptotic Variance for \foreignlanguage{american}{$\alpha$-pinene
model with} unknown initial conditions \label{tab:are_mse_a_pinene_unknown_ci}}
\end{table}

\par\end{center}

\subsection{Ricatti Equation \label{sub:Nonlinear-ODEs}}

%\subsubsection{Ricatti equation with a non-continuous vector field}

The true ODE is $\dot{\phi }=a\phi^{2}+c\sqrt{t}-d'\mathds{1}_{\left[T_{r};14\right]}$,
with $a^{*}=0.11$, $c^{*}=\text{0.09}$, $d^{*}=2$ and $\phi_{0}=-1$,
for $t\in\left[0,14\right]$. For all $\varphi$ in $C^{1}$ with
$\varphi(0)=\varphi(14)=0$, we have $<\phi,\dot{\varphi}>+a<\phi^{2},\varphi>+c<\sqrt{t},\varphi>-d'\left(\tilde{\varphi}(14)-\tilde{\varphi}(T_{r})\right)=0$
where $\tilde{\varphi}$ is the antiderivative of $\varphi$. \\
When $T_{r}$ is known, we use a cubic B-splines basis with 3 knots
at $T_{r}$, meaning that $\hat{\phi}$ can have a discontinuous derivative
at time $T_{r}$ (hence the curve estimation from noisy data is pretty
correct at $T_{r}$). The curve is mainly flat for $t\in\left[0,T_{r}\right]$
and after $T_{r}$, one can observe a linear behavior: 3 knots are
used to estimate the curve, and their positions are selected manually.\\
When $T_{r}$ is unknown, it is required to estimate $\theta=\left(a,c,d',T_{r}\right)$.
The OC is no more linear in parameters, but $\hat{\theta}^{OC}$ can
be computed by solving the general nonlinear program. The Two-Step
estimator fails to estimate $T_{r}$ because the derivative of the
solution is badly estimated when $T_{r}$ is unknown. OC estimators
still give reliable estimates as it uses only $\widehat{\phi}$ in
the criterion. Some care has to be taken for the knots selection because
of unknown $T_{r}$: when $n=200,\,400$ we use a uniform grid of
15 knots on $\left[0,\:14\right]$. For $n=50$, we have used 8 knots
uniformly located on $\left[0,\:14\right]$. Nevertheless, the nonparametric
estimates are too rough to obtaining any correct estimate $\hat{\theta}^{TS}$. 

Concerning NLS, we were not able to solve the optimization problem
and we cannot give Monte Carlo statistics for the evaluation of NLS.
NLS collapses in practice because the optimization problem is hard
(severely ill-posed problem). Indeed, the Levenberg-Marquardt algorithm
becomes very sensitive to initial conditions and gives different solutions
for very close starting values, even in the neighborhood of the true
value $\theta^{*}$. Moreover, we have to face with the problem of
explosion of the solutions during the optimization process. In particular,
this problem is very delicate because we have to chose $(a,c)$ so
that the (potential) explosion of the solution can be balanced by
a proper choice of $d'$ and $T_{r}$. Probably, NLS would benefit
from a specific optimization algorithm that could exploit the particular
properties of the ODE, but this is out of the scope of the paper. 

\begin{center}
\begin{table}[h]
\begin{centering}
\begin{tabular}{|c|c|c|c|c|c|c|}
\hline 
$\times10^{-2}$ & \multicolumn{3}{c|}{$MSE$} &  & \multicolumn{2}{c|}{$Tr\left(V\left(\hat{\theta}\right)\right)$}\tabularnewline
\hline 
\hline 
$(n,\sigma)$ & TS & OC & NLS &  & OC & NLS\tabularnewline
\hline 
$(400,0.2)$ & 0.18 & 0.27  & 0.58 &  & 1.76 & 0.10\tabularnewline
\hline 
$(400,0.4)$ & 0.78  & 1.21  & 0.94 &  & 2.56 & 0.38\tabularnewline
\hline 
$(200,0.2)$ & 0.33  & 0.87  & 0.57 &  & 2.85 & 0.25\tabularnewline
\hline 
$(200,0.4)$ & 1.12  & 2.69  & 1.12 &  & 5.64 & 0.98\tabularnewline
\hline 
$(50,0.2)$ & 1.03  & 1.30  & 1.54 &  & 4.70 & 1.00\tabularnewline
\hline 
$(50,0.4)$ & 3.80  & 4.43  & 3.94 &  & 8.89 & 4.08\tabularnewline
\hline 
\end{tabular}
\par\end{centering}

\caption{MSE , $Tr\left(V\left(\hat{\theta}\right)\right)$ for Parameter estimation
for Ricatti Equation with known $T_{r}$\label{tab:mse_are_ricatti_forcing_term}}
\end{table}

\par\end{center}

\begin{center}
\begin{table}[h]
\begin{centering}
\begin{tabular}{|c|c|c|c|c|}
\hline 
$\times10^{-2}$ & \multicolumn{1}{c|}{$MSE(\widehat{a})$} & \multicolumn{1}{c|}{$MSE(\widehat{c})$} & \multicolumn{1}{c|}{$MSE(\widehat{d'})$} & \multicolumn{1}{c|}{$MSE(\widehat{T}_{r})$}\tabularnewline
\hline 
\hline 
$(n,\sigma)$ & OC & OC & OC & OC\tabularnewline
\hline 
$(400,0.2)$ & 0.09 & 0.00 & 2.54 & 1.39\tabularnewline
\hline 
$(400,0.4)$ & 0.29 & 0.01 & 4.27 & 3.54\tabularnewline
\hline 
$(200,0.2)$ & 0.21 & 0.00  & 4.08 & 3.18\tabularnewline
\hline 
$(200,0.4)$ & 0.61 & 0.01 & 11.96 & 6.93\tabularnewline
\hline 
$(50,0.4)$ & 0.64 & 0.02 & 11.20 & 14.25\tabularnewline
\hline 
$(50,0.4)$ & 0.77 & 0.01 & 17.18 & 19.40\tabularnewline
\hline 
\end{tabular}
\par\end{centering}

\begin{centering}
\begin{tabular}{|c|c|c|}
\hline 
$\times10^{-2}$ & \multicolumn{1}{c|}{$MSE$} & \multicolumn{1}{c|}{$Tr\left(V(\hat{\theta})\right)$}\tabularnewline
\hline 
\hline 
$(n,\sigma)$ & OC & OC\tabularnewline
\hline 
$(400,0.2)$ & 4.01 & 3.97\tabularnewline
\hline 
$(400,0.4)$ & 8.11 & 8.02\tabularnewline
\hline 
$(200,0.2$) & 7.47 & 7.35\tabularnewline
\hline 
$(200,0.4)$ & 19.51 & 18.94\tabularnewline
\hline 
$(50,0.2$) & 26.10 & 5.14\tabularnewline
\hline 
$(50,0.4)$ & 37.36 & 9.49\tabularnewline
\hline 
\end{tabular}
\par\end{centering}

\caption{MSE, Sum Empirical Variance for Parameter estimation for Ricatti with
unknown $T_{r}$\label{tab:mse_are_ricatti_forcing_term_tr_unknown}}
\end{table}

\par\end{center}

\section{Real data analysis}

\subsection{Influenza virus growth and migration model}

We consider the ODE model introduced in Wu et. al \cite{Wu2011}
for the growth and migration of influenza virus-specific effector
CD8+ T cells, among lymph node ($T_{E}^{m}$), spleen ($T_{E}^{s}$),
and lung ($T_{E}^{l}$) of mice. After a model selection process,
it turns out that the following model 

\begin{equation}
\left\{ \begin{array}{l}
\frac{d}{dt}X_{1}=\rho_{m}D^{m}(t-\tau)-\gamma_{ms}\\
\frac{d}{dt}X_{2}=\rho_{s}D^{m}(t-\tau)-\gamma_{sl}+\gamma_{ms}e^{\left(X_{1}-X_{2}\right)}\\
\frac{d}{dt}X_{3}=\gamma_{sl}e^{\left(X_{2}-X_{3}\right)}-\delta_{l}
\end{array}\right.\label{eq:log_infl}
\end{equation}
is credible for representing the dynamics of the observations. Model
(\ref{eq:log_infl}) is written in log-scale (i.e with $X_{1}=log(T_{E}^{m})$,
$X_{2}=log(T_{E}^{s})$ and $X_{3}=log(T_{E}^{l})$), and the parameter
$\theta=(\rho_{m},\rho_{s},\delta_{l},\gamma_{ms},\gamma_{sl})^{T}$
has to be estimated from the data. The function $D$ and the delay
are known (estimated from the data).\\
The available data are the variables $T_{E}^{m}$, $T_{E}^{s}$ and
$T_{E}^{l}$ for six different subjects and are measured at times
$T=\left[0,\,4,\,5,\,6,\,7,\,8,\,9,\,10,\,11,\,12,\,14,\,24\right]$.
Following Wu et al., we stabilize the variance by a log transformation,
hence we consider directly the variables $X_{i},\: i=1,2,3$. We assume
that each subject share the same true parameter $\theta^{*}$ and
the same initial conditions: at each time point, we compute the mean
of the log-measurement (over the subjects) as pseudo-observations. 

We estimate $D^{m}$ with a spline smoother computed with cubic B-Splines
and GCV selection for the knots. As in Wu et al, the nonparametric
proxy is a regression spline $\widehat{X}=\left(\widehat{X_{1}},\widehat{X_{2}},\widehat{X_{3}}\right)$
defined on $\left[5,14\right]$; we do not consider earlier times
since the influenza specific CD8+ T cells are not produced before.
Since we have a small number of observations, the choice of the knots
for the cubic splines is done manually.

Nevertheless for the parameter estimation, we have tested several
estimates $\hat{X}$ (with different knots locations), and different
number of tests functions $L$:  we selected $L=3$
or $L=4$. The corresponding estimators are denoted $\hat{\theta}_{3}^{OC}$
and $\hat{\theta}_{4}^{OC}$. Moreover, in order to improve the accuracy , we have used a weighted version of the OC estimator, similar to the classical "Generalized Methods of Moments" (this procedure is detailed in section 5 of \emph{Supplementary Material I}). The quality of the estimator is evaluated
by the SSE:
\[
SSE=\sum_{s=1}^{6}\sum_{d=1}^{3}\sum_{i=1}^{N}\left(y_{i,d,s}-\phi_{d}(t_{i},\widehat{\theta},\widehat{X}(0))\right)^{2}
\]
where $y_{i,d,s}$ is the observation at time $t_{i}$ for the $s$-th
subject for the transformed variable $X_{d}$. As suggested in Wu
et al, we use the OC estimates as initial values for NLS estimation.
For both estimates, we obtain the same estimator which is then simply
denoted as $\hat{\theta}^{NLS}$.
We provide three different estimates $\hat{\theta}_{3}^{OC},\:\hat{\theta}_{4}^{OC}$
and $\hat{\theta}^{NLS}$; we mention also $\tilde{\theta}^{ref}$,
which is the estimate obtained in Wu et al  \cite{Wu2011}. 

\begin{center}
{\footnotesize }
\begin{table}[h]
\begin{centering}
{\footnotesize }%
\begin{tabular}{|c|c|c|c|c|}
\hline 
 & $\hat{\theta}_{3}^{OC}$ & $\hat{\theta}_{4}^{OC}$ & $\hat{\theta}^{NLS}$ & $\tilde{\theta}^{ref}$\tabularnewline
\hline 
\hline 
{\footnotesize $\rho_{m}$} & {\footnotesize 2.9e-5 } & {\footnotesize 2.7e-5 } & {\footnotesize 1.5e-5} & {\footnotesize $1.6e-5$}\tabularnewline
\hline 
{\footnotesize $\rho_{s}$} & {\footnotesize 4.1e-5} & {\footnotesize 4.7e-5} & {\footnotesize 4.1e-5 } & {\footnotesize $4.5e-5$}\tabularnewline
\hline 
{\footnotesize $\delta_{l}$} & {\footnotesize 2.0} & {\footnotesize 3.4} & {\footnotesize 3.7  } & {\footnotesize $3.96$}\tabularnewline
\hline 
{\footnotesize $\gamma_{ms}$} & {\footnotesize 0.39} & {\footnotesize 0.35} & {\footnotesize 0.15} & {\footnotesize $0.157$}\tabularnewline
\hline 
{\footnotesize $\gamma_{sl}$} & {\footnotesize 0.72 } & {\footnotesize 0.81 } & {\footnotesize 0.47} & {\footnotesize $0.49$}\tabularnewline
\hline 
{\footnotesize RMSE} & {\footnotesize $13.5$} & {\footnotesize $13.9$} & {\footnotesize $9.0$} & {\footnotesize $9.5$}\tabularnewline
\hline 
\end{tabular}
\par\end{centering}{\footnotesize \par}

\centering{}{\footnotesize }%
\begin{tabular}{|c|c|c|c|c|c|c|}
\hline 
 & \multicolumn{2}{c|}{$\hat{\theta}_{3}^{OC}$} & \multicolumn{2}{c|}{$\hat{\theta}_{4}^{OC}$} & \multicolumn{2}{c|}{$\hat{\theta}^{NLS}$}\tabularnewline
\hline 
\hline 
 & Low. Bound & Up. Bound & Low. Bound & Up. Bound & Low. Bound & Up. Bound\tabularnewline
\hline 
{\footnotesize $\rho_{m}$} & {\footnotesize 2.1e-5} & {\footnotesize 3.7e-5} & {\footnotesize 1.9e-5} & {\footnotesize 3.4e-5} & {\footnotesize 0.7e-0.5 } & {\footnotesize 2.4e-0.5}\tabularnewline
\hline 
{\footnotesize $\rho_{s}$} & {\footnotesize 0.7e-5} & {\footnotesize 7.4e-5} & {\footnotesize 0.9e-5} & {\footnotesize 8.4e-5} & {\footnotesize 3.4e-0.5} & {\footnotesize 4.8e-0.5}\tabularnewline
\hline 
{\footnotesize $\delta_{l}$} & {\footnotesize -1.11} & {\footnotesize 5.21} & {\footnotesize -0.28} & {\footnotesize 7.21} & {\footnotesize 2.59} & {\footnotesize 4.93}\tabularnewline
\hline 
{\footnotesize $\gamma_{ms}$} & {\footnotesize 0.27} & {\footnotesize 0.50} & {\footnotesize 0.24 } & {\footnotesize 0.46} & {\footnotesize 0.03} & {\footnotesize 0.26}\tabularnewline
\hline 
{\footnotesize $\gamma_{sl}$} & {\footnotesize -0.10} & {\footnotesize 1.55} & {\footnotesize -0.14} & {\footnotesize 1.76} & {\footnotesize 0.39} & {\footnotesize 0.55}\tabularnewline
\hline 
\end{tabular}{\footnotesize \caption{\label{tab:Estimation-resultsWu}Estimates, RMSE and the 95\% confidence
intervals for different $L$ and estimators. }
}
\end{table}

\par\end{center}{\footnotesize \par}

\begin{center}
\begin{figure}[h]
\begin{centering}
\includegraphics[scale=0.5]{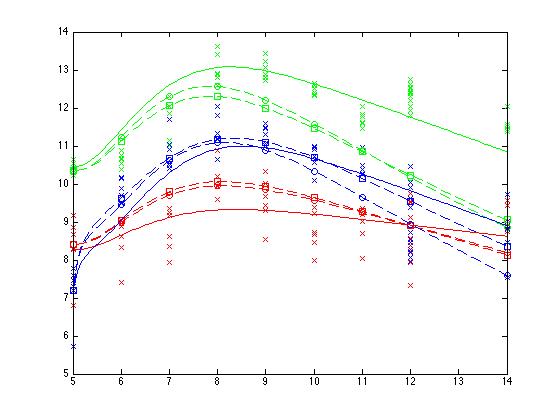}\caption{Influenza model, Estimated curves for $X_{1}$ (red), $X_{2}$ (green), $X_{3}$ (blue);
$\times$: observations, $\square$: solution for $\hat{\theta}_{1}^{OC}$
, $\circ$: solution for $\hat{\theta}_{2}^{OC}$, solid line: solution
with $\hat{\theta}^{NLS}$.\label{fig:curve_influenza}}

\par\end{centering}

\end{figure}

\par\end{center}

\begin{center}
\begin{figure}[h]
\centering{}\includegraphics[scale=0.5]{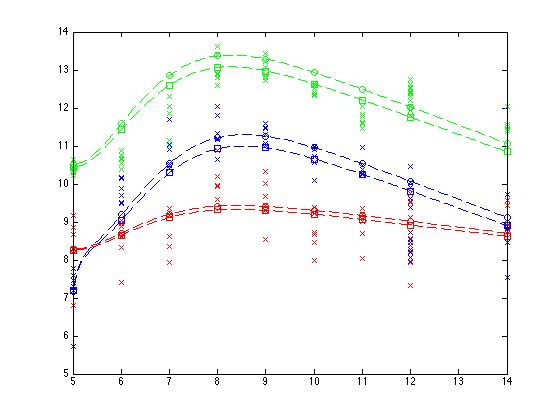}\caption{Influenza model, Estimated curves for $X_{1}$ (red), $X_{2}$ (green), $X_{3}$ (blue);
$\square$ solution obtained with OC+NLS, $\circ$ solution obtained
with $\tilde{\theta}^{ref}$. \label{fig:curve_influenza_oc_vs_wu}}
\end{figure}

\par\end{center}

\subsection{Blowfly model}

The Delay Differential Equation (\ref{eq:NicholsonDDE}) was proposed
by Gurney et al \cite{Gurney1980} to model the dynamics of a population
of blowflies, from the Nicholson's blowfly data \cite{Nicholson1957}.
These data consists of 350 counts taken every two days during between
day $40=T_{0}$ and day $315=T_{1}$. As Gurney did, we take $\tau=14.8$
days and our aim is to estimate $\theta=\left(P,N_{0},\delta\right)$.
The orthogonal conditions derived from the weak form is $\forall\varphi\in C_{c}^{1}\left(\left]a,b\right[\right)$, 
\[
\int_{a}^{b}N(u)\dot{\varphi}(u)du+P\int_{a-\tau}^{b-\tau}N(u)e^{-\frac{N(u)}{N_{0}}}\varphi(u+\tau)du-\delta\int_{a}^{b}N(u)\varphi(u)du=0
\]
where $\left[a,b\right]$ has to be chosen such that: $\left[a,b\right]$
,$\left[a-\tau,b-\tau\right]\subset\left[T_{0},T_{1}\right]$. Due
to a change in the dynamics, we have used only the first 180 observations,
see \cite{Ellner2002}. For the nonparametric estimation, we have
used $42$ knots located between $t=40$ and $t=220$. Preliminary
tests and comparisons suggests to use the sine basis for the test
function $\varphi_{\ell}$, and we use $2\leq L\leq15$. A simulation is given in figure \ref{fig:curve_Blowfly}

\begin{center}
{\footnotesize }
\begin{table*}[h]
\begin{centering}
{\footnotesize }%
\begin{tabular}{|c|c|c|c|}
\hline 
 & {\footnotesize $L=11$} & {\footnotesize $L=9$} & {\footnotesize $L=12$}\tabularnewline
\hline 
\hline 
{\footnotesize $P$} & {\small 7.81} & {\small 7.52} & {\small 7.91}\tabularnewline
\hline 
{\footnotesize $N_{0}$} & {\small 381.8} & {\small 385.9} & {\small 377.7}\tabularnewline
\hline 
{\footnotesize $\delta$} & {\small 0.154} & {\small 0.153} & {\small 0.154}\tabularnewline
\hline 
{\small RSSE} & {\small 1.7136e+03} & {\small 1.7557e+03} & {\small 1.7990e+03}\tabularnewline
\hline 
\end{tabular}
\par\end{centering}{\footnotesize \par}

\centering{}{\footnotesize }%
\begin{tabular}{|c|c|c|c|c|c|c|}
\hline 
 & \multicolumn{2}{c|}{{\footnotesize $L=11$}} & \multicolumn{2}{c|}{{\footnotesize $L=9$}} & \multicolumn{2}{c|}{{\footnotesize $L=12$}}\tabularnewline
\hline 
\hline 
{\footnotesize O.C} & Low. Bound & Up. Bound & Low. Bound & Up. Bound & Low. Bound & Up. Bound\tabularnewline
\hline 
{\footnotesize $P$} & {\footnotesize 5.80 } & {\footnotesize 9.81} & {\footnotesize 5.64} & {\footnotesize 9.40} & {\footnotesize 5.0416 } & {\footnotesize 10.77}\tabularnewline
\hline 
{\footnotesize $N_{0}$} & {\footnotesize 303.62 } & {\footnotesize 459.94} & {\footnotesize 306.59} & {\footnotesize 465.38} & {\footnotesize 289.36} & {\footnotesize 465.98}\tabularnewline
\hline 
{\footnotesize $\delta$} & {\footnotesize 0.10} & {\footnotesize 0.20} & {\footnotesize 0.11} & {\footnotesize 0.19} & {\footnotesize 0.10 } & {\footnotesize 0.20}\tabularnewline
\hline 
\end{tabular}{\footnotesize \caption{\label{tab:est_nicholson}Estimates, RSSE and 95\% confidence intervals
for different $L$}
}
\end{table*}

\par\end{center}

\begin{center}
\begin{figure}[h]
\centering{}\includegraphics[scale=0.5]{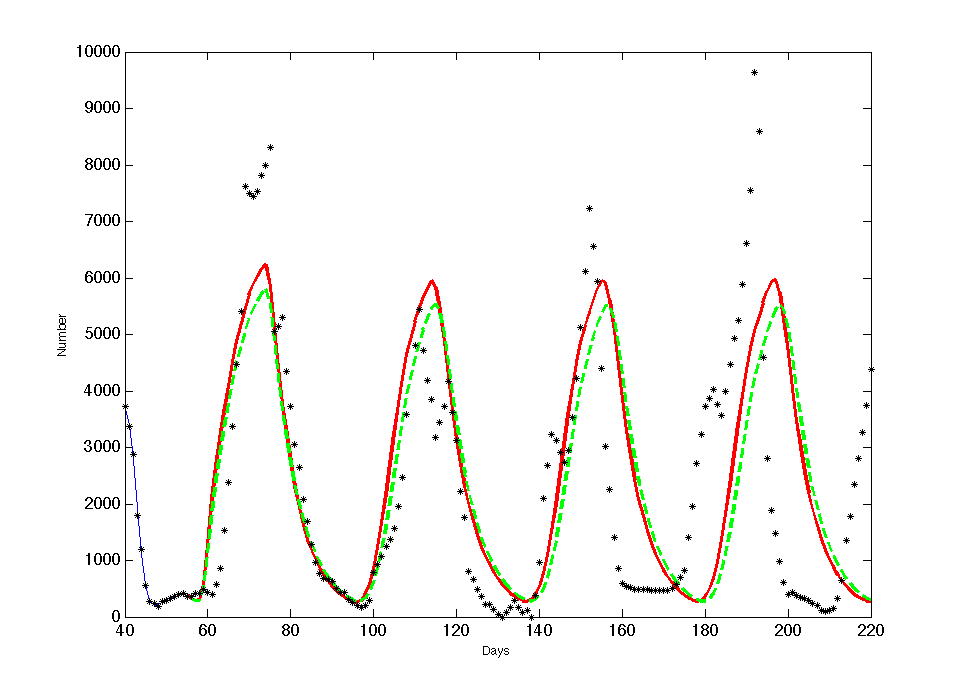}\caption{Solution $N$ of the Nicholson's DDE simulated with the OC estimator (computed with $L=11$ conditions - continuous red line). The NLS solution is given by the dashed green curve. The initial function is estimated between day $40$ and $55$ and the simulation starts after day 55. Drift between data and simulations comes from a chaotic behavior and uncertainty in initial condition (and parameters)\label{fig:curve_Blowfly}}
\end{figure}

\par\end{center}

%However due to the non-linearity of the vector field w.r.t parameters and state
%variables, the variance is not well estimated and the IRWOC failed
%to improve estimation (and does not converge). Hence, we give the
%OC estimators $\hat{\theta}_{9}^{OC},\,\hat{\theta}_{11}^{OC},\,\hat{\theta}_{12}^{OC}$
%computed with uniform weight i.e without using IRWOC. 

\section{Discussion}
Among the simulated models we considered ($\alpha$-pinene, Ricatti), the NLS estimator is often the best estimator in the asymptotic case (and small noise case) in terms of MSE for the parameters. Nevertheless, in some complex case such as unknown initial conditions for $\alpha$-pinene (with small sample size or high noise level), or Ricatti equation (with known or unknown change point $T_r$), then TS and OC can offer better statistical performances. 
The $\alpha$-pinene model shows the interest of using information on the boundaries in OC (as introduced in section 2.1). Moreover, simulations show that OC can improve on classical TS although it uses only (partial information) about (weak) derivatives. The fact that the NLS can be caught up, even in the very favorable case of a closed-form solution and starting values (for NLS optimization) close to the true parameter indicates that the introduction of Functions Moments offers a competitive estimator to the direct classical for complex case. 
In the latter case of Ricatti, the TS approaches is uniformly better than NLS, whereas OC is not systematically better than NLS. 
Ricatti Equation is striking, as it shows that good proxies $\hat{\phi}$ gives a lot of information: when $T_r$ is known, the reconstruction of the solution and its derivative is excellent, which gives a clear advantage to the plain TS. Nevertheless, when $T_r$ is unknown the derivative estimation is of poor quality around $T_r$, and the TS estimator is unstable and  cannot be computed. The same situation occurs for NLS, because of some lack of identifiability and dramatic changes in derivative estimation which makes the   optimization algorithms inefficient. 
For the influenza dataset analysis, the two OC estimators give correct parameter estimates from real and sparse data (the simulated  ODE have a correct qualitative behavior). When used as starting for NLS, both estimates give the same NLS estimator, which improves (obviously) the SSE and still gives an estimator closer to the estimates given Wu et al (and same qualitative behavior for the solution). We consider the (self-)consistency of the OC estimates as an indication for reliability of the OC approach. More generally, OC can be used for initializing a NLS estimator, which is often a critical problem in nonlinear regression. In our case, we found a slightly better estimate (for RSS) w.r.t the original paper by Wu et al.
For the Delay Differential Equation modeling the blowfly dataset, we insist on the ease of implementation of the method, that avoids the semiparametric estimation of the initial condition. Moreover it provides an estimate close to the posterior mean obtained by ABC: $P^{ABC}=7.39$, $N_0^{ABC}=365.03$ and $\delta^{ABC}=0.15$. With a varying number of Orthogonal Conditional, we can assess the self-consistency of our estimate. Moreover, the posterior mean is always in the $95\%$ confidence set computed for OC.

\subsection*{Acknowledgments}

This work was funded by two ANR projects GD2S (ANR-05-MMSA-0013-01) and ODESSA (ANR-09-SYSC-009-01)
and received a partial support from the Analysis of Object Oriented Data program (2010-2011) in Statistical and Applied Mathematical 
Sciences Institute (SAMSI), USA.

\bibliographystyle{plain}
\bibliography{biblio_weak}

\end{document}